\documentclass[journal]{IEEEtran}
\IEEEoverridecommandlockouts

\usepackage{sty/style}

\newtheorem{lem}[prop]{Lemma}
\newtheorem{cor}[prop]{Corollary}
\newtheorem{prty}{Property}

\newcommand*{\fancyreflemlabelprefix}{lem}
\frefformat{vario}{\fancyreflemlabelprefix}{Lem.~#1}
\newcommand*{\fancyrefcorlabelprefix}{cor}
\frefformat{vario}{\fancyrefcorlabelprefix}{Cor.~#1}
\frefformat{vario}{\fancyrefapplabelprefix}{Appendix~#1}
\newcommand*{\fancyrefprtylabelprefix}{prty}
\frefformat{vario}{\fancyrefprtylabelprefix}{Property~#1}
\frefformat{vario}{\fancyreftablabelprefix}{Table~#1}
\usepackage{booktabs}

\usetikzlibrary{fit,backgrounds}  
\definecolor{crabred}{HTML}{BE1831}

\pgfplotsset{
  boundStyle/.style = {black,     dotted,         line width=1.0pt},
  biaStyle/.style   = {black!55,  solid,          line width=1.0pt,
                       mark=triangle*, mark size=2.2pt},
  unbStyle/.style   = {black!55,  densely dashed, line width=1.0pt,
                       mark=diamond*,   mark size=2.4pt},
  coilStyle/.style  = {red,       solid,          line width=1.3pt,
                       mark=*,         mark size=2.0pt},
  appStyle/.style   = {blue,      densely dashed, line width=1.0pt,
                       mark=square*,    mark size=1.8pt},
  ceilStyle/.style  = {black,     dashed,         line width=1.0pt,
                       mark=x,         mark size=2.6pt},
  iddStyle/.style   = {green_120, dashdotted,     line width=1.0pt,
                       mark=pentagon*, mark size=2.0pt},
  floodMod/.style   = {dotted, line width=0.8pt, mark options={fill=white},
                       mark size=1.6pt},
}

\DeclareRobustCommand{\varnodesym}{\tikz[baseline={([yshift=-0.55ex]vn.center)}]{%
  \node[draw, circle, inner sep=0.15pt, minimum size=2.3mm, line width=0.5pt, font=\tiny] (vn) {$\scriptscriptstyle=$};}}

\DeclareRobustCommand{\cchknodesym}{\tikz[baseline={([yshift=-0.55ex]dn.center)}]{%
  \node[draw, rectangle, inner sep=0.15pt, minimum size=2.1mm, line width=0.5pt, font=\tiny, fill=blue!12] (dn) {$\scriptscriptstyle+$};}}

\newcommand{\herm}{^\He}      \newcommand{\trans}{^\T}
\newcommand{\setO}{\mathcal{O}}
\newcommand{\setA}{\mathcal{A}}  
\newcommand{\setE}{\mathcal{E}}

\newcommand{\setS}{\mathcal{S}}
\newcommand{\Reals}{\mathbb{R}}
\newcommand{\Complex}{\mathbb{C}}
\newcommand{\Nzero}{N_0}
\newcommand{\Es}{E_{\up{s}}}
\newcommand{\MT}{M_{\up{T}}}
\newcommand{\MR}{M_{\up{R}}}
\newcommand{\Ncw}{N_{\up{c}}}
\newcommand{\Ninfo}{N_{\up{i}}}
\newcommand{\SNR}{\textit{SNR}}
\newcommand{\eqdef}{\triangleq}
\newcommand{\e}{\mathrm{e}}
\newcommand{\simdir}{gmi_sim/out}  
\newcommand{\simdirerr}{idd_sim/out}  

\newcommand{\boxplusop}{\boxplus}

\allowdisplaybreaks %%%% THIS THING ALLOWS EQUATIONS TO BREAK THE PAGE

\begin{document}

%%%%%%%%%%%%%%%%%%%%%%%%%%%%%%%%%%%%%%%%%%%%%%%%%%%%%%%%%%%
%% Title, authors, and thanks
%%%%%%%%%%%%%%%%%%%%%%%%%%%%%%%%%%%%%%%%%%%%%%%%%%%%%%%%%%%

% Working title; still not perfect
\title{Mitigating the BICM Loss in MIMO Receivers\\ with Correlation-Informed Logits}

\author{\IEEEauthorblockN{Christoph Studer}\thanks{CS is with the Department of Information Technology and Electrical Engineering at ETH Zurich, Switzerland; email: studer@ethz.ch}%
\thanks{This work was funded in part by armasuisse and by the Swiss State Secretariat for Education, Research, and Innovation (SERI) under the SwissChips initiative. This work also received support from the Swiss National Science Foundation (SNSF) through the Project ``Ubiquitous Large InTelligent ArRAys (ULTRA)'' under Grant 219710 and the HORIZON-JU-SNS-2024-STREAM-C ``X-TREME 6'' Project under Grant 101192681.}%
\thanks{The author thanks A.~Kumar for useful comments on the paper. The author used Anthropic Claude Opus 4.8 during preparation of this manuscript for language and grammar editing, and as an auxiliary tool for checking mathematical derivations and MATLAB code. Claude was not used to generate scientific results or conclusions. All AI-assisted suggestions and outputs were independently evaluated and, where applicable, verified by the author, who takes full responsibility for the content of the manuscript.}
}

\maketitle

%%%%%%%%%%%%%%%%%%%%%%%%%%%%%%%%%%%%%%%%%%%%%%%%%%%%%%%%%%%
%% Abstract
%%%%%%%%%%%%%%%%%%%%%%%%%%%%%%%%%%%%%%%%%%%%%%%%%%%%%%%%%%%

\begin{abstract}
Bit-interleaved coded modulation (BICM) is widely used in multiple-input multiple-output (MIMO) wireless systems. 
However, separating MIMO data detection from channel decoding and only exchanging per-bit soft information discards correlation information among the bits within each transmit vector.
We show that the resulting information loss limits the performance of practical one-shot receivers. Even though iterative detection and decoding (IDD) can recover most of this loss, IDD significantly increases receiver complexity and latency.
In order to reduce this loss, we propose COrrelation-Informed Logits (COIL), a soft-output format that combines per-bit soft information with pairwise coupling logits for correlated bits.
We show how to incorporate these coupling logits into a conventional belief-propagation-based channel decoder, which enables correlation-aware decoding in a one-shot receiver without IDD.
We evaluate COIL in a MIMO-BICM system with a conventional low-density parity-check code. 
For a $\bf 4\boldsymbol\times 4$ MIMO orthogonal frequency-division multiplexing (OFDM) system, a one-shot linear minimum mean-square error (LMMSE) receiver with COIL outperforms a conventional one-shot unbiased LMMSE receiver by 1.3\,dB and operates within 0.8\,dB of four-iteration IDD, while requiring only a single pass through the data detector.
\end{abstract}


%%%%%%%%%%%%%%%%%%%%%%%%%%%%%%%%%%%%%%%%%%%%%%%%%%%%%%%%%%%
%% Introduction
%%%%%%%%%%%%%%%%%%%%%%%%%%%%%%%%%%%%%%%%%%%%%%%%%%%%%%%%%%%

\section{Introduction}
\IEEEPARstart{B}{it}-interleaved coded modulation (BICM)~\cite{zehavi92,caire98,martinez09} is the de facto standard in modern multiple-input multiple-output (MIMO) wireless systems as it achieves excellent error-rate performance at manageable complexity~\cite{caire98,hochwald03,seethaler04}.
The basic idea is as follows: The transmitter encodes information bits with a modern forward error-correction code, such as a low-density parity-check (LDPC) code~\cite{gallager62,richardson01,richardson08}, and spreads the resulting coded bits (typically after interleaving) over space, frequency, and/or time.
The receiver then operates in two stages. First, a soft-output MIMO data detector produces per-coded-bit reliability information, commonly called soft outputs. Second, a soft-input channel decoder uses this soft information to compute estimates of the transmitted information bits.

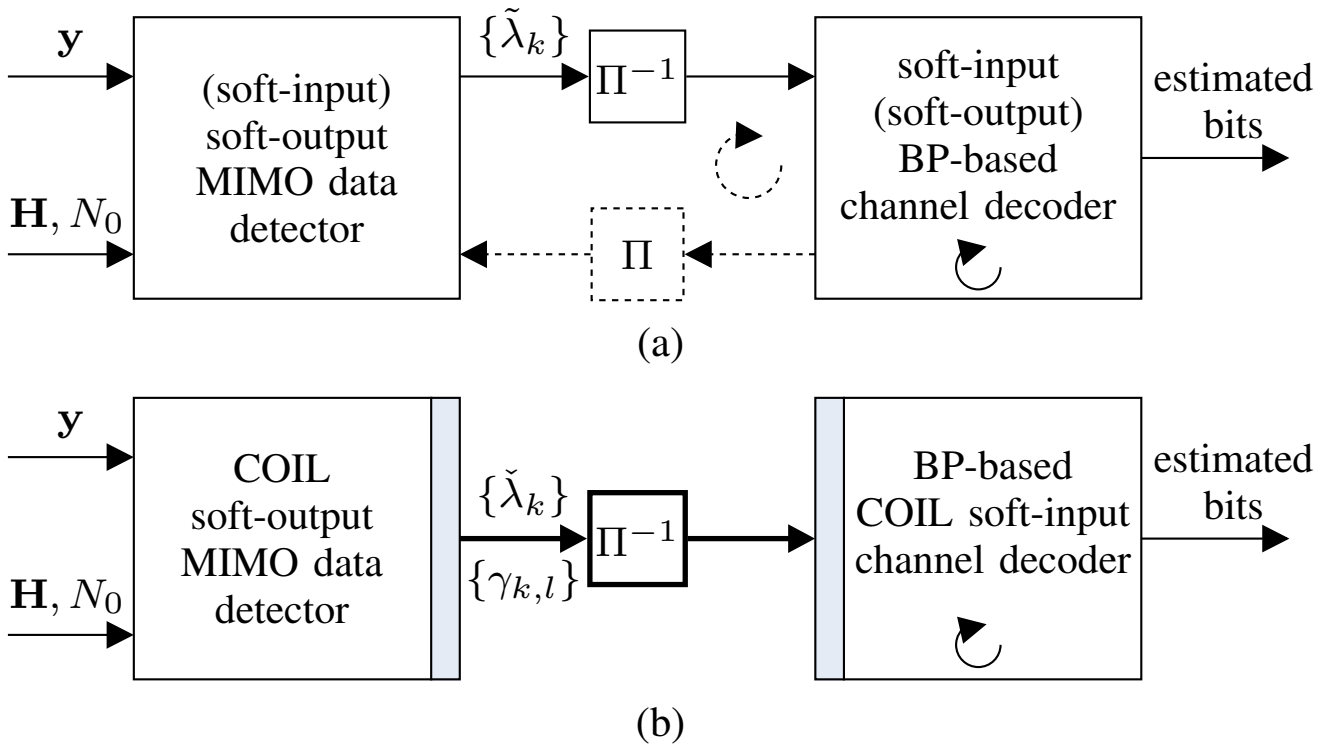
\begin{figure}[tp]
\centering
%%% standard receiver
\subfloat[\label{fig:rxconv}]{%
\begin{tikzpicture}[
scale=0.98, transform shape,
font=\footnotesize,
blk/.style={draw, minimum width=2.20cm, minimum height=1.90cm, align=center, inner sep=2pt},
sml/.style={draw, minimum width=0.62cm, minimum height=0.62cm, inner sep=1pt},
lbl/.style={font=\footnotesize, inner xsep=1.5pt, inner ysep=3.5pt}, >={Triangle[width=1.8mm,length=1.8mm]},
loop/.pic={\draw[->] (0.15,0) arc[start angle=0, end angle=-292, radius=0.15];},
dloop/.pic={\draw[->,dash pattern=on 1.2pt off 1.2pt] (0.22,0) arc[start angle=0, end angle=-296, radius=0.22];},
]
\node[blk] (det) at (1.95,0) {(soft-input)\\soft-output\\MIMO data\\detector};
\node[sml] (dil) at (4.25,0.55) {$\Pi^{-1}$};
\node[sml, dash pattern=on 1.2pt off 1.2pt] (fil) at (4.25,-0.65) {$\Pi$};
\node[blk] (dec) at (6.55,0) {};
\node[align=center] at ([yshift=0.16cm]dec.center) {soft-input\\(soft-output)\\BP-based\\channel decoder};
\draw[->] (0.00,0.55) -- (0.85,0.55) node[lbl,pos=0.50,above] {$\vec{y}$};
\draw[->] (0.00,-0.65) -- (0.85,-0.65) node[lbl,pos=0.50,above] {$\mat{H},\Nzero$\,};
\draw[->] (3.05,0.55) -- (dil.west) node[lbl,midway,above] {$\{\tilde{\lambda}_k\}$};
\draw[->] (dil.east) -- (5.45,0.55);
\draw[dash pattern=on 1.2pt off 1.2pt,->] (5.45,-0.65) -- (fil.east);
\draw[dash pattern=on 1.2pt off 1.2pt,->] (fil.west) -- (3.05,-0.65);
\pic at (5.00,-0.05) {dloop};
\pic at ([yshift=0.22cm]dec.south) {loop};  
\draw[->] (7.65,0) -- (8.65,0) node[lbl,pos=0.60,above,align=center] {\,estimated\\\,bits};
\end{tikzpicture}}
\\[0.15cm]
%%% COIL receiver
\subfloat[\label{fig:rxcoil}]{%
\begin{tikzpicture}[
scale=0.98, transform shape,
font=\footnotesize,
blk/.style={draw, minimum width=2.20cm, minimum height=1.90cm, align=center, inner sep=2pt},
sml/.style={draw, minimum width=0.62cm, minimum height=0.62cm, inner sep=1pt},
lbl/.style={font=\footnotesize, inner xsep=1.5pt, inner ysep=3.5pt}, >={Triangle[width=1.8mm,length=1.8mm]},
hatch/.style={fill=blue!12},
loop/.pic={\draw[->] (0.15,0) arc[start angle=0, end angle=-292, radius=0.15];},
]
\node[blk] (cdet) at (1.95,0) {};
\node[align=center] at ([xshift=-0.10cm]cdet.center) {COIL\\soft-output\\MIMO data\\detector};
\node[sml, line width=1.0pt] (cdil) at (4.25,0) {$\Pi^{-1}$};
\node[blk] (cdec) at (6.55,0) {};
\node[align=center] at ([xshift=0.10cm,yshift=0.18cm]cdec.center) {BP-based\\COIL soft-input\\channel decoder};
\begin{scope}[on background layer]
\fill[hatch] ([xshift=-0.20cm]cdet.north east) rectangle (cdet.south east);
\fill[hatch] (cdec.north west) rectangle ([xshift=0.20cm]cdec.south west);
\end{scope}
\draw ([xshift=-0.20cm]cdet.north east) -- ([xshift=-0.20cm]cdet.south east);
\draw ([xshift=0.20cm]cdec.north west) -- ([xshift=0.20cm]cdec.south west);
\draw[->] (0.00,0.55) -- (0.85,0.55) node[lbl,pos=0.50,above] {$\vec{y}$};
\draw[->] (0.00,-0.65) -- (0.85,-0.65) node[lbl,pos=0.50,above] {$\mat{H},\Nzero$\,};
\draw[->,line width=1.0pt] (3.05,0) -- (cdil.west) node[lbl,midway,above] {$\{\check{\lambda}_k\}$} node[lbl,midway,below] {$\{\gamma_{k,l}\}$};
\draw[->,line width=1.0pt] (cdil.east) -- (5.45,0);
\pic at ([yshift=0.24cm]cdec.south) {loop};
\draw[->] (7.65,0) -- (8.65,0) node[lbl,pos=0.60,above,align=center] {\,estimated\\\,bits};
\end{tikzpicture}}
%%%
\caption{MIMO-BICM receiver architectures.
(a)~Conventional receiver: the data detector generates soft outputs $\{\tilde{\lambda}_k\}$, which the deinterleaver $\Pi^{-1}$ passes to the channel decoder.
(b)~Proposed COIL receiver: the data detector additionally computes coupling logits $\{\gamma_{k,l}\}$ (shaded left), which are processed by the channel decoder using a statically augmented factor graph (shaded right). Circular arrows mark iterations; the dashed path in~(a) illustrates the outer iteration of an IDD receiver which COIL avoids.}
\label{fig:receivers}
\end{figure}

In order to reduce complexity and latency, most MIMO-BICM receivers (i) are \emph{one-shot} (see \fref{fig:rxconv}), i.e., they run the data detector and the channel decoder only once, and (ii) use \emph{equalization-based} data detectors, which first estimate the transmitted symbols and then compute per-coded-bit soft information~\cite{butler04,seethaler04}.
In many situations, more sophisticated soft-output data detectors and/or iterative detection and decoding (IDD) can significantly improve the error-rate performance of MIMO-BICM systems~\cite{hochwald03,studer10,studer11,wiesmayr22} (see the dashed feedback path in \fref{fig:rxconv}).
IDD iterates between a soft-input soft-output MIMO data detector and a soft-input soft-output channel decoder. These (outer) iterations increase complexity and latency, complicate scheduling of the two stages, and make efficient hardware implementation difficult~\cite{studer09}.

A key component of MIMO-BICM is the \emph{interface} between the two stages through which the data detector passes per-coded-bit soft information to the channel decoder.
This interface enables us to develop, optimize, and implement both stages separately, which is essential for efficient hardware implementations~\cite{burg06,haene06,haene08,studer08,studer11}. However, such a per-bit interface throws away crucial information contained in the received signals.

%%%%%%%%%%%%%%%%%%%%%%%%%%%%%%%%%%%%%%%%%%%%%%%%%%%%%%%%%%%
\subsection{Contributions}

We propose a one-shot MIMO-BICM receiver architecture that recovers a large part of the BICM loss using \emph{COrrelation-Informed Logits} (COIL); see \fref{fig:rxcoil}.
We start with a fundamental question: \emph{where exactly does a one-shot MIMO-BICM receiver lose performance?}
We answer this question by analyzing the detector--decoder interface using a vector-level generalized mutual information (GMI) measure that gives an achievable rate for the Gaussian metric produced by an equalization-based MIMO data detector.
Our analysis reveals that the rate loss is caused primarily by the interface between the data detector and the channel decoder, rather than by the detector's suboptimality: per-bit soft outputs discard crucial correlation among the bits mapped to the same transmit vector.
This insight motivates COIL, a soft-output format in which an equalization-based MIMO data detector not only reports per-coded-bit soft outputs but also pairwise \emph{coupling logits} that convey correlation information.
We first analyze how much correlation information COIL provides. We then show that message-passing-based channel decoders can process COIL soft outputs by statically augmenting their factor graph with one additional degree-three check node per coupled bit pair---the decoding algorithm is left unchanged.
With COIL, all iterations remain within the channel decoder, and no outer iterations over the MIMO data detector are required, which reduces complexity and latency.
As a proof of concept, we simulate a MIMO orthogonal frequency-division multiplexing (OFDM) system with a conventional LDPC code.
Our results demonstrate that a one-shot linear minimum mean-square error (LMMSE)-based receiver that outputs COIL outperforms conventional one-shot receivers and approaches IDD receivers at a fraction of their complexity and~latency.
Since COIL changes only the receiver, it can be deployed in any system that builds upon MIMO-BICM without any modifications to the communication standard.

%%%
\subsection{Relevant Prior Art}
\label{sec:priorart}

\subsubsection{BICM}
Zehavi proposed the architecture now known as BICM in~\cite{zehavi92}. Caire, Taricco, and Biglieri subsequently established its information-theoretic foundations and showed that the per-bit metric interface loses rate relative to coded modulation~\cite{caire98}. Reference~\cite{martinez09} later reinterpreted the bit-metric interface as mismatched decoding, for which the GMI~\cite{kaplan93,ganti00} is the standard achievable-rate measure; references~\cite{guillen08,scarlett20} discuss BICM and mismatched decoding in detail.
In contrast, we ask how much information is contained in a MIMO data detector's output \emph{before} it is reduced to per-bit metrics. Furthermore, we show that the BICM loss corresponds to the conditional total correlation~\cite{watanabe60} of the coded bits within one transmit vector, given the receive vector.
COIL recovers the part of this total correlation that can be captured by a maximum-weight spanning tree. The spanning tree's vertices (or nodes) correspond to the coded bits of one transmit vector, and each edge is associated with one coupling logit that characterizes the dependence between the connected bit~pair.

\subsubsection{Non-Binary Codes}
COIL is not the only approach to improve upon the per-bit interface of BICM. For example, non-binary LDPC codes over higher-order Galois fields~\cite{davey98} enable the decoder to consider dependence among groups of bits, and such codes are known to improve performance~\cite{davey98}.
However, non-binary LDPC codes fix the grouping at code-design time and require specialized decoder algorithms, which are difficult to implement in hardware, as discussed in, e.g.,~\cite{declercq07,voicila10}.
In contrast, COIL retains the binary code and the decoding algorithm, and lets the data detector select the coupled bit pairs for each transmit vector from the received signal. Thus, COIL is compatible with the binary LDPC codes of modern wireless standards, whereas non-binary codes are not.

\subsubsection{MIMO-BICM}
MIMO-BICM is at the heart of the physical layer of IEEE~802.11n Wi-Fi~\cite{dot11n09} as well as 3GPP long-term evolution (LTE)~\cite{lte36211} and 5G new radio (NR)~\cite{nr38211}, and it is the standard setting of the MIMO soft-output detection literature~\cite{hochwald03,seethaler04,haene08,studer08,studer11}.
Existing rate-based comparisons assess soft-output data detectors through the per-bit mutual information (MI) between each transmitted bit and its detector output, summed into the system-capacity metric~\cite{fertl12}.
In contrast, we measure the information contained in the \emph{summary} produced by an equalization-based data detector: its estimate (the mean) and the full error covariance matrix. This analysis identifies the discarded covariance information as the dominant loss mechanism and motivates COIL.

\subsubsection{IDD}
IDD builds upon the turbo principle~\cite{berrou93,hagenauer02} and originated in the turbo equalization and multiuser detection literature~\cite{douillard95,wang99,tuchler02}.
Iterative receivers for MIMO-BICM systems were proposed in, e.g.,~\cite{tonello00,sellathurai02}, and reference~\cite{hochwald03} showed that IDD with a list sphere decoder achieves near-capacity performance; subsequent work refined the algorithms and implemented them in hardware~\cite{studer09,studer11,wiesmayr22}. 
All of these architectures capture inter-bit correlation by re-running the data detector with updated a priori information.
In contrast, we capture inter-bit correlation \emph{once} and incorporate this information directly into the factor graph of a belief-propagation (BP)-based channel decoder. 
Our architecture (cf.\ \fref{fig:rxcoil}) is therefore \emph{one-shot}: it runs the MIMO data detector once, avoiding the complexity, latency, and scheduling difficulties of the IDD outer loop~\cite{studer09}.

\subsubsection{Joint detection-decoding factor graphs}
Factor graphs~\cite{kschischang01,loeliger04} have been used to design iterative receivers that fuse detection and decoding, e.g.,~\cite{boutros02,kurkoski02}.
For example, reference~\cite{kurkoski02} augments a message-passing LDPC decoder with the trellis of a partial-response channel and iterates between the two.
Similarly, message-passing-based data detectors~\cite{cespedes14,jeon15,ma17,rangan19,arai25,clausius25} exchange messages between a linear estimation module and a nonlinear denoising module; these messages typically consist of per-symbol means and scalar variances and do not convey inter-bit correlation.
In contrast, we explicitly pass inter-bit correlation information from the data detector to the channel decoder, which processes this information with a conventional BP-based decoding algorithm by augmenting its factor graph.

\subsubsection{Tree-structured approximations}
Chow--Liu trees~\cite{chowliu68} and the total correlation~\cite{watanabe60} are classical tools for tree-structured approximations of joint distributions and for measuring the information loss of such approximations, respectively; tree-structured approximations have also been used in receivers before.
For example, the MIMO data detector of~\cite{goldberger11} constructs a Chow--Liu tree over the transmitted symbols from a Gaussian relaxation and performs BP on that tree to detect uncoded data; the LDPC decoder of~\cite{salamanca13} imposes a tree structure on pairs of uncertain variables during decoding and adds pairwise constraints that capture information missed by BP.
In contrast, COIL constructs a tree for the coded bits of a transmit vector and passes the pairwise correlation information along its edges to the channel decoder as a correlation-aware soft-output format.
The MIMO data detector computes this pairwise correlation information and passes it to the channel decoder as a small number of scalars. The channel decoder then processes these scalars using the usual BP updates.

%%%%%%%%%%%%%%%%%%%%%%%%%%%%%%%%%%%%%%%%%%%%%%%%%%%%%%%%%%%
\subsection{Notation}

Matrices and column vectors are written as boldface uppercase and lowercase letters, respectively. For a matrix~$\mat{A}$, the transpose is $\mat{A}\trans$, the conjugate transpose is $\mat{A}\herm$, the entry in the $\ell$th row and $k$th column is $[\mat{A}]_{\ell,k}$, and $\mathrm{diag}(\mat{A})$ zeros all off-diagonal entries of $\mat{A}$.
The identity matrix is~$\mat{I}$ and the all-zeros matrix (or vector) is $\mat{0}$; their dimensions will be clear from context.
For a vector~$\vec{a}$, the $k$th entry is $a_k$, the $\ell_2$-norm is $\|\vec{a}\|_2$, the real part is $\Re(\vec{a})$, and the imaginary part is $\Im(\vec{a})$.
The real-valued decomposition of a complex-valued vector $\vec{a}\in\Complex^{M}$ is $\vec{a}_\Reals\eqdef[\Re(\vec{a})\trans,
\Im(\vec{a})\trans]\trans\in\Reals^{2M}$.
Sets are written with uppercase calligraphic letters and the cardinality of the set $\setS$ is $|\setS|$.
The ceiling operator $\lceil a\rceil$ returns the smallest integer that is greater than or equal to $a\in\Reals$.
Expectation is denoted by $\Ex[\cdot]$; the averaged quantity will be clear from context. Probabilities are denoted by $\Pr[\cdot]$.
The distribution of a circularly-symmetric complex Gaussian random vector with covariance matrix $\mat{C}$ is denoted by~$\mathcal{CN}(\vec{0},\mat{C})$.
Bits take values in $\{+1,-1\}$: we map the (logical) bit values $0$ and $1$ to $+1$ and $-1$, respectively. 
The entropy of a discrete random quantity $a$ is $H(a)$, the conditional entropy of~$a$ given $b$ is $H(a\,|\,b)$, the MI between~$a$ and $b$ is $I(a;b)$, and the Kullback--Leibler (KL) divergence between two probability mass functions (PMFs) $p$ and $q$ on the same finite set~$\setS$ is $\mathrm{KL}(p\,\|\,q)\eqdef\sum_{a\in\setS}p(a)\log_2(p(a)/q(a))$; all of these are measured in bits.
We use hats for quantities computed by the data detector (e.g.,~$\hat{\vec{s}}$ and $\hat{\mat{C}}$), tildes for soft outputs that approximate exact logits (e.g., $\tilde{\lambda}$), and checks for field-corrected soft outputs (e.g., $\check{\lambda}$).

%%%%%%%%%%%%%%%%%%%%%%%%%%%%%%%%%%%%%%%%%%%%%%%%%%%%%%%%%%%
\subsection{Paper Outline}

The rest of the paper is organized as follows. \fref{sec:prereq} introduces MIMO-BICM and equalization-based soft-output MIMO data detection.
\fref{sec:loss} compares several GMI metrics and identifies that one-shot MIMO-BICM loses information by discarding inter-bit correlation.
\fref{sec:coil} introduces the COIL soft-output format and the COIL-based data detector; \fref{sec:receiver} details the COIL-based channel decoder with a BP-based LDPC decoder as an example.
\fref{sec:results} reports simulation results.
\fref{sec:limitations} summarizes limitations and open problems.
We conclude in \fref{sec:conclusions}.
The appendix contains all proofs.

%%%%%%%%%%%%%%%%%%%%%%%%%%%%%%%%%%%%%%%%%%%%%%%%%%%%%%%%%%%
%% Prerequisites
%%%%%%%%%%%%%%%%%%%%%%%%%%%%%%%%%%%%%%%%%%%%%%%%%%%%%%%%%%%

\section{Prerequisites}
\label{sec:prereq}

We start by introducing the MIMO-BICM system model. We then formalize the class of equalization-based soft-output MIMO data detectors considered throughout the paper.

%-----------------------------------------------------------
\subsection{MIMO-BICM Overview}
\label{sec:model}

We consider a spatial-multiplexing system with $\MT$ transmit streams (layers) and $\MR$ receive antennas that transmits over~$K$ resource blocks (e.g., subcarriers or time steps).

\subsubsection{Transmitter}
At the transmitter, a channel encoder maps a block of $\Ninfo$ information bits to a codeword containing $\Ncw$ coded bits.
The coded bits are (typically) interleaved and partitioned into $K=\lceil\Ncw/N\rceil$ subvectors $\vec{x}=[x_1,\dots,x_N]\trans\in\{\pm1\}^{N}$ of $N\eqdef\MT Q$ bits each (if $N$ does not divide $\Ncw$, the last subvector is padded with zero bits), and every subvector is mapped to a transmit vector $\vec{s}\in\setO^{\MT}$ that occupies one resource block.\footnote{We focus on a point-to-point scenario with coding across antennas; an extension to the multi-user MIMO uplink is straightforward.}
Here, $\setO\subset\Complex$ is the constellation set (e.g., 16-QAM) with $Q\eqdef\log_2|\setO|$ bits per symbol. 
We use $n(k)\in\{1,\dots,\MT\}$ and $b(k)\in\{1,\dots,Q\}$ to denote the stream and the label index, respectively, of the $k$th bit; put simply, $x_k$ is the $b(k)$th label bit of the symbol~$s_{n(k)}$.
We use $\setO_{b}^{\,x}\subset\setO$ to denote the set of constellation symbols in~$\setO$ whose $b$th label bit corresponds to~$x\in\{\pm1\}$. 
We assume (i) Gray labeling\footnote{\label{fn:labeling}Gray labeling is standard in MIMO-BICM. The IDD literature, however, has shown that other labelings (e.g., anti-Gray) can perform better~\cite{tenbrink98,chindapol01}.} and (ii) that the coded bits within one transmit vector are uniform and i.i.d., which is a standard BICM assumption~\cite{caire98}.
The latter implies that the transmit vectors $\vec{s}$ are uniformly distributed over the set of transmit vectors~$\setO^{\MT}$.

\subsubsection{MIMO Channel}
We model the channel matrices of the~$K$ resource blocks as a stationary and ergodic random sequence; this implies that all resource blocks are statistically identical and we omit the resource-block index.\footnote{Our rate expressions only depend on the per-resource-block marginal distribution; independence across resource blocks is not required.}
We model the received vector $\vec{y}\in\Complex^{\MR}$ of one resource block as follows:
\begin{align}
  \vec{y} = \mat{H}\vec{s} + \vec{z}.
  \label{eq:iorel}
\end{align}
Here, $\mat{H}\in\Complex^{\MR\times\MT}$ is the channel matrix (we assume frequency-flat channels within one resource block) and $\vec{z}\sim\mathcal{CN}(\vec{0},\Nzero\mat{I})$ models noise that is independent of $\vec{s}$ and $\mat{H}$ and i.i.d.\ across resource blocks (we assume $\Nzero\,{\in}\,\Reals_{>0}$).
For simplicity, we assume that~$\mat{H}$ and $\Nzero$ are known perfectly\footnote{We need this assumption only when computing exact logits and $I(\vec{s};\vec{y})$. All GMI-based rates used here remain achievable for imperfect channel knowledge, i.e., a detector using an estimated $\hat{\mat{H}}$ and/or $\hat{N}_0$ simply computes a more mismatched summary, whose achievable rate the GMI measures.} at the receiver;
we define $\Es\eqdef\Ex[|s_n|^2]$, $n=1,\ldots,\MT$, where the expectation is over the constellation, and we assume $\Ex[\vec{s}\vec{s}\herm]=\Es\mat{I}$.
Throughout the paper, we use the average receive signal-to-noise ratio (SNR) defined as $\SNR\eqdef\MT\Es/\Nzero$.

\subsubsection{Receiver}
At the receiver (cf.\ \fref{fig:rxconv}), a soft-output MIMO data detector computes reliability information for the~$N$ coded bits of every transmit vector. Optimal inference about the transmit vector $\vec{s}\in\setO^{\MT}$ (and hence about its bits) depends on the receive vector only through the posterior
\begin{align}
  p(\vec{s}\,|\,\vec{y}) \propto
  \e^{-\|\vec{y}-\mat{H}\vec{s}\|_2^2/\Nzero}
  \label{eq:posterior}
\end{align}
over~$\vec{s}\in\setO^{\MT}$.
Practical soft-output data detectors for MIMO-BICM systems compute per-bit reliability information that approximates the per-bit marginals $\Pr[x_k\,|\,\vec{y}]$ of \fref{eq:posterior}, as detailed in \fref{sec:eqdet}. The reliability information from all $K$ resource blocks is typically deinterleaved and passed to a soft-input channel decoder, which estimates the transmitted information bits.
In one-shot BICM, this pipeline is executed exactly once (cf.\ the solid feed-forward path in \fref{fig:rxconv}); in conventional IDD, the decoder feeds (extrinsic or intrinsic) soft information back to the soft-input soft-output MIMO data detector and the pipeline iterates (cf.\ the dashed feedback path in \fref{fig:rxconv}).

%-----------------------------------------------------------
\subsection{Equalization-Based Soft-Output MIMO Data Detection}
\label{sec:eqdet}

Many practical soft-output MIMO data detectors are equalization-based: they first form an estimate of the transmit vector and then convert this estimate into per-bit soft information under a Gaussian error model. We now formalize this two-stage structure. The defining property is the format of the summary generated by the data detector; the data detector itself can be linear or nonlinear. 

\begin{defi}[Equalization-Based Soft-Output MIMO Data Detection] \label{defi:eqdet}
An equalization-based soft-output MIMO data detector proceeds in two stages:
\begin{enumerate}
\item \emph{Equalization:} a (linear or nonlinear) equalizer $f$ computes the estimate $\hat{\vec{s}} = f(\vec{y},\mat{H},\Nzero)$ of the transmit vector, along with the covariance matrix $\hat{\mat{C}}\eqdef\Ex[\vec{e}\vec{e}\herm]$ and the pseudo-covariance matrix $\hat{\mat{K}}\eqdef\Ex[\vec{e}\vec{e}\trans]$ of the residual error $\vec{e}\eqdef\vec{s}-\hat{\vec{s}}$, which we assume to have zero mean; here, expectation is over the transmit and noise vectors for a given channel matrix $\mat{H}$.\footnote{For many nonlinear equalizers, the matrices~$\hat{\mat{C}}$ and $\hat{\mat{K}}$ are estimated rather than computed exactly; our results do not require them to be exact.}
\item \emph{Soft-output computation:} the residual error $\vec{e}$ is modeled as Gaussian with the covariance matrix~$\hat{\mat{C}}$ and pseudo-covariance matrix~$\hat{\mat{K}}$ from the first stage; all soft outputs are extracted from the resulting Gaussian model and the constellation set $\setO$.
\end{enumerate}
We refer to the tuple $\{\hat{\vec{s}},\hat{\mat{C}},\hat{\mat{K}}\}$ as the
data detector's \emph{summary}. 
\end{defi}

A general (improper) Gaussian error model with $\hat{\mat{K}}\neq\mat{0}$ is handled through the real-valued decomposition, i.e., the real-valued error vector $\vec{e}_\Reals$ has the real-valued covariance matrix
\begin{align}
  \hat{\mat{C}}_\Reals = \textstyle \frac{1}{2}
  \begin{bmatrix}
    \Re(\hat{\mat{C}}+\hat{\mat{K}}) & \Im(\hat{\mat{K}}-\hat{\mat{C}}) \\
    \Im(\hat{\mat{K}}+\hat{\mat{C}}) & \Re(\hat{\mat{C}}-\hat{\mat{K}})
  \end{bmatrix}\!,
  \label{eq:realdecomp}
\end{align}
so that the tuple $\{\hat{\mat{C}},\hat{\mat{K}}\}$ is equivalent to the single matrix $\hat{\mat{C}}_\Reals\in\Reals^{2\MT\times2\MT}$. 
For the rest of the paper, we assume that the error~$\vec{e}$ is proper, i.e., $\hat{\mat{K}}=\mat{0}$.
For one-dimensional constellations (e.g., BPSK or PAM), the error is generally improper. For best performance, the pseudo-covariance should therefore be included through the representation in~\fref{eq:realdecomp}.

\begin{rem}
\fref{defi:eqdet} covers linear data detectors, such as the well-known zero-forcing (ZF) and soft-output LMMSE data detectors~\cite{seethaler04}, as well as nonlinear data detectors, such as LAMA~\cite{jeon15}, OAMP/VAMP~\cite{ma17,rangan19}, and expectation-propagation-based detectors~\cite{cespedes14}, all of which produce a mean and an error covariance (typically only its diagonal).
\end{rem}

The goal of Stage~(ii) is to approximate the exact logits of the $N$ bits of a transmit vector. Since the distinction between exact and approximate
reliability information on the coded bits is central to this paper, we use the following definitions.

\begin{defi}[Logits and Soft Outputs] \label{defi:logit}
The \emph{logit}\footnote{In the literature, logits are often called log-likelihood ratios (LLRs).} of the $k$th bit of a transmit vector is the exact log-odds under the true posterior~\fref{eq:posterior}:
\begin{align}
  \lambda_k \eqdef \ln\!\left(\frac{\Pr[x_k=+1\,|\,\vec{y}]}{\Pr[x_k=-1\,|\,\vec{y}]}\right)\!.
  \label{eq:exactlogit}
\end{align}
We call approximations of~\fref{eq:exactlogit} computed by practical data detectors \emph{soft outputs} and mark them with a tilde: $\tilde{\lambda}_k$.
\end{defi}

Direct evaluation of~\fref{eq:exactlogit} requires marginalizing the posterior~\fref{eq:posterior} over all $|\setO|^{\MT}$ transmit vectors. Equalization-based data detectors approximate this computation to reduce complexity.
A data detector that evaluates~\fref{eq:exactlogit} exactly is called an a posteriori probability (APP) data detector; we will use it as a benchmark.

The common approach to compute soft outputs from the summary $\{\hat{\vec{s}},\hat{\mat{C}}\}$ proceeds \emph{per stream}: the $n$th entry of $\hat{\vec{s}}$ is modeled as a Gaussian observation of $s_n$ with error variance~$[\hat{\mat{C}}]_{n,n}$, which induces the per-stream symbol metric
\begin{align}
  \phi_n(a) \eqdef \e^{-|\hat s_n-a|^2/[\hat{\mat{C}}]_{n,n}},
  \label{eq:symbolmetric}
\end{align}
with $a\in\setO$. The soft output of bit $k$ is then obtained by marginalization over the constellation set as
\begin{align}
  \tilde{\lambda}_k \eqdef \ln\!\left(
  \frac{\sum_{a\in\setO_{b(k)}^{+1}} \phi_{n(k)}(a)}
       {\sum_{a\in\setO_{b(k)}^{-1}} \phi_{n(k)}(a)}  \right)\!.
  \label{eq:logit}
\end{align}
We emphasize that soft-output computation via~\fref{eq:symbolmetric} and~\fref{eq:logit} effectively replaces the summary $\{\hat{\vec{s}},\hat{\mat{C}}\}$ of \fref{defi:eqdet} by the \emph{reduced} summary $\{\hat{\vec{s}},\mathrm{diag}(\hat{\mat{C}})\}$: only the diagonal of the error covariance matrix enters the metric. 
In many cases, the data detector has already computed the off-diagonal entries of $\hat{\mat{C}}$ (or could compute them at low complexity), but discards them before computing the soft outputs.

%-----------------------------------------------------------
\subsection{Soft-Output LMMSE and ZF MIMO Data Detection}
\label{sec:lmmse}

As a concrete instance of \fref{defi:eqdet}, we consider the widely used soft-output LMMSE data detector~\cite{seethaler04}. For simplicity, we focus on the biased variant.
With the Gram matrix $\mat{G}\eqdef\mat{H}\herm\mat{H}$ and the regularized Gram matrix $\mat{A}\eqdef\mat{G}+\tfrac{\Nzero}{\Es}\mat{I}$, the equalization stage computes the following summary:
\begin{align} \label{eq:lmmse}
  \hat{\vec{s}}_{\up{L}} \eqdef \mat{A}^{-1}\mat{H}\herm\vec{y}
  \quad \text{and} \quad
  \hat{\mat{C}}_{\up{L}} \eqdef \Nzero\mat{A}^{-1}.
\end{align}
The covariance matrix in~\fref{eq:lmmse} is the \emph{exact} error covariance of the biased LMMSE estimate.
The soft-output stage then uses only the diagonal entries of $\hat{\mat{C}}_{\up{L}}$ to form the per-stream symbol metric as in~\fref{eq:symbolmetric} and the soft outputs as in~\fref{eq:logit}.
We note that the full error covariance matrix $\hat{\mat{C}}_{\up{L}}$ in~\fref{eq:lmmse} might be available as a by-product of computing~$\hat{\vec{s}}_{\up{L}}$. Most hardware implementations, however, only compute the diagonal entries; see, e.g., the hardware designs in~\cite{haene06,studer11}.

In the low-noise limit, the LMMSE equalization stage reduces to soft-output ZF equalization~\cite{butler04}. Provided that $\mat{G}$ has full rank, for $\Nzero/\Es\to0$, the summary of~\fref{eq:lmmse} is
\begin{align}  \label{eq:zf}
  \hat{\vec{s}}_{\up{ZF}} \eqdef \mat{G}^{-1}\mat{H}\herm\vec{y}
  \quad \text{and} \quad
  \hat{\mat{C}}_{\up{ZF}} \eqdef \Nzero\mat{G}^{-1}.
\end{align}
Soft outputs are computed from the ZF summary~\fref{eq:zf} exactly as for LMMSE, i.e., via \fref{eq:symbolmetric} and \fref{eq:logit}.
In our context, there exists one key difference between the two equalizers:

\begin{rem}\label{rem:zfexact}
For any linear equalizer $\hat{\vec{s}}=\mat{W}\vec{y}$, the residual error corresponds to $\vec{e}=(\mat{I}-\mat{W}\mat{H})\vec{s}-\mat{W}\vec{z}$, which is independent of the transmit vector if and only if $\mat{W}\mat{H}=\mat{I}$. The ZF equalizer $\mat{W}=\mat{G}^{-1}\mat{H}\herm$ satisfies this equality and its error $\vec{e}=-\mat{G}^{-1}\mat{H}\herm\vec{z}\sim\mathcal{CN}(\vec{0},\Nzero\mat{G}^{-1})$ is Gaussian and proper ($\hat{\mat{K}}=\mat{0}$). 
Thus, the Gaussian error model of \fref{defi:eqdet} is \emph{exact}. For every linear equalizer that is not a left inverse of~$\mat{H}$---such as the LMMSE equalizer---a signal-dependent term, which is non-Gaussian for discrete constellations, remains. As a result, the Gaussian error model is only an approximation.
\end{rem}

%-----------------------------------------------------------
\subsection{Soft-Input Channel Decoder}
\label{sec:decoder}

The soft-input channel decoder takes the (typically deinterleaved) soft outputs of the $\Ncw$ coded bits, collected from all resource blocks, and computes estimates of the transmitted information bits. Two aspects of this per-bit interface are important for our analysis.
First, most practical channel decoders treat their inputs as exact, i.e., the decoding rule assumes that the soft outputs are exact logits as in~\fref{eq:exactlogit}.
When the data detector provides soft outputs, i.e., approximate logits, as most practical detectors do, the decoder performs \emph{mismatched decoding}~\cite{ganti00,martinez09}. A suitable achievable-rate measure must therefore account for such a mismatch.
Second, the channel decoder scores codewords using per-bit soft inputs only; we therefore restrict the considered soft-input channel decoders to the following class. 

\begin{defi}[Bit-Metric Decoding] \label{defi:bmd}
A soft-input channel decoder performs \emph{bit-metric decoding} (BMD)~\cite{martinez09} if it scores every codeword $\vec{c}\in\{\pm1\}^{\Ncw}$ by the bit-wise product metric
\begin{align}
  q_{\up{bit}}(\vec{c}) = \prod_{k=1}^{\Ncw} q_k(c_k)
  \quad \text{with} \quad
  q_k(c_k) \eqdef
  \frac{1}{1+\e^{-c_k\tilde{\lambda}_k}}.
  \label{eq:productmetric}
\end{align}
Here, $q_k$ is the PMF whose log-odds are given by the soft output $\tilde{\lambda}_k$ of the $k$th coded bit~$c_k$.
\end{defi}

\begin{rem} \label{rem:agnostic}
The decoder class of~\fref{eq:productmetric} includes conventional LDPC, turbo, polar, and convolutional channel decoders, since all of them (in their standard form) process per-bit soft inputs.
\end{rem}

%%%%%%%%%%%%%%%%%%%%%%%%%%%%%%%%%%%%%%%%%%%%%%%%%%%%%%%%%%%
%% Where do we lose information? CORRELATION!
%%%%%%%%%%%%%%%%%%%%%%%%%%%%%%%%%%%%%%%%%%%%%%%%%%%%%%%%%%%

\section{Where Does One-Shot MIMO-BICM Lose~Information?}
\label{sec:loss}

We now measure the information rate that is lost by one-shot MIMO-BICM. To this end, we first review the bit-level GMI~\cite{kaplan93,ganti00,martinez09}, which is the standard measure for soft-output quality. 
We then introduce a vector-level GMI measure that applies to every equalization-based MIMO data detector in the sense of \fref{defi:eqdet}. We use this measure to show that the dominant loss mechanism is the discarded correlation among the bits of a transmit vector, rather than the detector's inherent suboptimality.
In what follows, all rates are in bits per transmit vector and are averaged over the channel realizations.\footnote{Since the random channel matrix~$\mat{H}$ is known at the receiver, all posteriors, conditional entropies, and MIs involving~$\vec{y}$ are implicitly conditioned on~$\mat{H}$. For example, we write $p(\vec{s}\,|\,\vec{y})$, $H(x_k\,|\,\vec{y})$, $I(x_k;x_l\,|\,\vec{y})$, and $I(\vec{s};\vec{y})$ as shorthand for the corresponding quantities additionally conditioned on~$\mat{H}$; expectations over~$\vec{y}$ also include the channel realizations.}

%%%
\subsection{Decoding Metrics and Achievable Rate}
\label{sec:metricgmi}
The three rates used in this paper are the GMI of the same channel for three different decoding metrics. 
We now define a general metric and its rate, and then invoke this definition for the three individual decoding metrics. 

\begin{defi}[Decoding Metric and Its GMI]
\label{defi:gmi}
Let $\setA$ denote the input alphabet of one resource block, either the set $\setO^{\MT}$ of transmit vectors or its bit labeling $\{\pm1\}^N$. In both cases, we have~$N=\log_2|\setA|$, and we write $\vec{x}\in\setA$ for the transmitted (bit) vector of one resource block.
A \emph{decoding metric} is any strictly positive function $q$ on $\setA$ that the receiver computes from $\vec{y}$, $\mat{H}$, and $\Nzero$.
The associated rate (in bits per transmit vector) is the GMI~\cite{kaplan93,ganti00,martinez09}
\begin{align} \label{eq:gmi}
R(q) \eqdef N + \max_{\tau\geq0}\ \Ex\!\left[ \log_2\!\left(\frac{q(\vec{x})^\tau}{\sum_{\vec{x}'\in\setA}q(\vec{x}')^\tau} \right)\right]\!.
\end{align}
Here, $\tau\geq0$ is the scaling parameter of mismatched decoding and the expectation is over the transmitted data, the channel realizations, and the noise.
We prove in \fref{app:gmi} that every rate below~\fref{eq:gmi} is achievable.
\end{defi}

The next lemma provides a tool for evaluating the three decoding metrics used in this paper.
A \emph{decoding metric}~$q$ is not necessarily a PMF. For every $\tau\geq0$, however, we can define a PMF on the input alphabet $\vec{x}\in\setA$ by raising the metric to the power $\tau\geq0$ and normalizing it as
\begin{align} \label{eq:pmfthingie}
\bar{q}_\tau(\vec{x}) \eqdef \frac{q(\vec{x})^\tau}{\sum_{\vec{x}'\in\setA}q(\vec{x}')^\tau}.
\end{align}
The following lemma states that the rate of a decoding metric is the MI $I(\vec{s};\vec{y})$ reduced by the KL divergence between the true posterior in~\fref{eq:posterior} and the PMF in \fref{eq:pmfthingie}. 
For a given metric, the rate loss is the smallest expected KL divergence from the true posterior to the normalized metrics obtained by varying~$\tau$. 
Although the proof in \fref{app:divergence} is elementary, we will invoke the following lemma several times.

\begin{lem}[GMI as Posterior Divergence] \label{lem:divergence}
For a decoding metric $q$ and $\tau\geq0$, let $\bar{q}_\tau$ be the associated PMF in \fref{eq:pmfthingie}. 
Then, the following statements hold:
\begin{enumerate}
\item The rate~\fref{eq:gmi} for a fixed $\tau$, denoted by $R_\tau(q)$, satisfies
\begin{align} \label{eq:divergence}
R_\tau(q) = I(\vec{s};\vec{y}) - \Ex\!\left[\mathrm{KL}\big( p(\cdot\,|\,\vec{y})\,\big\|\,\bar{q}_\tau\big)\right]\!,
\end{align}
so that $R(q)=I(\vec{s};\vec{y})-\min_{\tau\geq0}\Ex[\mathrm{KL}(p(\cdot\,|\,\vec{y})\|\bar{q}_\tau)]$.
This implies $R(q)\leq I(\vec{s};\vec{y})$, with equality if and only if some choice of~$\tau\geq0$ makes $\bar{q}_\tau$ equal to the posterior, i.e., $q^\tau\propto p(\cdot\,|\,\vec{y})$, for all receive vectors~$\vec{y}$ (except for a set of probability zero).
\item Let $q$ be a PMF. If, for all receive vectors~$\vec{y}$ (except for a set of probability zero), the quantity $\ln(q)$ has the same expectation under the posterior as under $q$ itself, i.e., $\Ex_{p(\cdot|\vec{y})}[\ln(q)] = \Ex_{q}[\ln(q)]$, then $\tau=1$ maximizes~\fref{eq:gmi} and $R(q)=I(\vec{s};\vec{y})-\Ex[\mathrm{KL}(p(\cdot\,|\,\vec{y})\|q)]$.
\end{enumerate}
\end{lem}

Statement~(ii) is a calibration condition that is easy to check in practice.
Suppose that $\ln(q)$ is a sum of terms, each of which depends on only a small group of bits, e.g., on one bit or one bit pair at once.
The expectation of $\ln(q)$ then depends only on the marginals of these groups of bits.
Consequently, if $q$ and the posterior have the same marginals for these groups of bits, then the two averages in statement~(ii) coincide and $\tau=1$ is optimal.
We will use statement~(ii) for the exact per-bit metric in \fref{prop:bicmloss} and the exact tree metric in \fref{prop:oracletree}; it is the reason why the optimal scaling parameter is $\tau^\star=1$ for both metrics.

%-----------------------------------------------------------
\subsection{Bit-Level Generalized Mutual Information}
\label{sec:bitgmi}

Consider a bit-metric decoder that treats the soft outputs $\tilde{\lambda}_k$ of the data detector as exact logits~$\lambda_k$. 
Since these soft outputs are, for most practical detectors, mismatched (cf.\ \fref{sec:decoder}), the appropriate achievable rate is the GMI~\cite{kaplan93,ganti00,martinez09}.

\begin{defi}[Bit-Metric GMI]\label{defi:bitgmi}
Let a data detector report soft outputs $\{\tilde{\lambda}_1,\dots,\tilde{\lambda}_N\}$ for one transmit vector, with any amount of model mismatch. 
Let $R_{\up{bit}}\eqdef R(q_{\up{bit}})$ be the BMD rate (measured in bits per transmit vector), i.e., \fref{eq:gmi} evaluated for the product metric~\fref{eq:productmetric}. 
This BMD rate corresponds to
\begin{align}\label{eq:bitgmi}
R_{\up{bit}} = N + \max_{\tau\geq0} \>\sum_{k=1}^{N} \> \Ex\!\left[ \log_2 \!\left( \frac{q_k(x_k)^\tau} {\sum_{x\in\{\pm1\}} q_k(x)^\tau} \right) \right]\!.
\end{align}
\end{defi}

Inserting the per-bit metrics~\fref{eq:productmetric} into~\fref{eq:bitgmi} and rearranging terms yields the well-known expression~\cite{caire98,martinez09}
\begin{align}
  R_{\up{bit}} = \max_{\tau\geq0}\ \sum_{k=1}^{N}
  \left( 1 - \Ex\!\left[\log_2\!\big(1+\e^{-\tau x_k\tilde{\lambda}_k}\big)\right] \right)\!.
  \label{eq:bitgmiexplicit}
\end{align}
Since the expectation in~\fref{eq:bitgmi} includes channel realizations, $R_{\up{bit}}$ is an ergodic rate, achievable by coding across resource blocks.

\begin{rem} \label{rem:fertl}
Reference~\cite{fertl12} assessed soft-output MIMO data detectors using the histogram-estimated MI between transmitted bits and their corresponding soft outputs and discussed its GMI interpretation for suitably corrected soft outputs.
The MI measures what optimally post-processed soft outputs would achieve and is invariant to invertible miscalibrations. 
In contrast, the GMI in~\fref{eq:bitgmi} measures what the channel decoder actually uses and is affected by miscalibration. Since practical channel decoders usually trust their inputs, we focus solely on GMI-based metrics.
\end{rem}

\begin{rem} \label{rem:sparam}
The parameter~$\tau^\star$ that maximizes~\fref{eq:bitgmi} measures the confidence of the soft outputs: $0<\tau^\star<1$ indicates overconfident soft outputs, $\tau^\star>1$ indicates underconfident soft outputs, and $\tau^\star\approx1$ indicates well-calibrated soft outputs.
For exact logits, i.e., $\tilde{\lambda}_k=\lambda_k$, $\tau^\star=1$ and each bit's contribution to~\fref{eq:bitgmi} equals the per-bit mutual information $I(x_k;\vec{y})$, so that $R_{\up{bit}}=\sum_{k=1}^{N}I(x_k;\vec{y})$; this can be obtained by invoking \fref{lem:divergence}(ii) and is shown as part of the proof of \fref{prop:bicmloss}.
\end{rem}

The bit-metric GMI in \fref{eq:bitgmi} measures the quality of the soft outputs of the $N$ coded bits of one transmit vector. 
This metric alone cannot answer our key question: \emph{how much information does the data-detector output contain before we reduce it to per-bit soft outputs?} To answer this question, we now propose an information measure that operates at the vector level.

%%%
\subsection{Vector-Level Generalized Mutual Information}
\label{sec:vecgmi}

We propose an achievable rate measure that can be evaluated for \emph{any} equalization-based MIMO data detector in the sense of \fref{defi:eqdet}, i.e.,
for any data detector that reports the summary~$\{\hat{\vec{s}},\hat{\mat{C}},\hat{\mat{K}}\}$ or its equivalent real-valued decomposition~$\{\hat{\vec{s}}_\Reals,\hat{\mat{C}}_\Reals\}$ via~\fref{eq:realdecomp}. 
Specifically, we determine the achievable rate of a decoder that uses the Gaussian metric defined by the detector's summary. This decoder therefore follows the statistical model assumed in the second stage of \fref{defi:eqdet}.

\begin{defi}[Vector-Level GMI]
\label{defi:vecgmi}
Let a data detector report the summary $\{\hat{\vec{s}}_\Reals,\hat{\mat{C}}_\Reals\}$ with a full-rank covariance matrix~$\hat{\mat{C}}_\Reals$ and any amount of model mismatch. Define the Gaussian decoding metric over the constellation set as
\begin{align}
  q_{\up{vec}}(\vec{s}\,|\,\{\hat{\vec{s}}_\Reals,\hat{\mat{C}}_\Reals\}) \eqdef
  \e^{-\frac{1}{2}(\vec{s}_{\Reals}-\hat{\vec{s}}_{\Reals})\trans
  \hat{\mat{C}}_{\Reals}^{-1}(\vec{s}_{\Reals}-\hat{\vec{s}}_{\Reals})},
  \label{eq:vecmetric}
\end{align}
where we write $q_{\up{vec}}(\vec{s})$ as a shortcut for $q_{\up{vec}}(\vec{s}\,|\,\{\hat{\vec{s}}_\Reals,\hat{\mat{C}}_\Reals\})$. 
Let $R_{\up{vec}}\eqdef R(q_{\up{vec}})$ be the vector-metric decoding (VMD)
rate. Then, we have 
\begin{align}
  R_{\up{vec}} = N + \max_{\tau\geq0}\>
  \Ex\!\left[
    \log_2 \! \left( \frac{q_{\up{vec}}(\vec{s})^\tau}
    {\sum_{\vec{s}'\in\setO^{\MT}} q_{\up{vec}}(\vec{s}')^\tau} \right)
    \right]\!.
  \label{eq:vecgmi}
\end{align}
\end{defi}

The VMD rate is the GMI of the Gaussian metric in~\fref{eq:vecmetric} induced by the detector's summary.
Since this rate can be compared across all equalization-based MIMO data detectors, we use it as a benchmark between $I(\vec{s};\vec{y})$ and the rate of the conventional bit metric in~\fref{eq:productmetric}.
The VMD rate measures the summary \emph{before} the per-bit reduction of \fref{sec:eqdet}; the BMD rate measures it \emph{after} this reduction. 
The nonnegative quantity $I(\vec{s};\vec{y})-R_{\up{vec}}$ measures the rate loss of the Gaussian vector metric~\fref{eq:vecmetric}. In this paper, $R_{\up{vec}}-R_{\up{bit}}$ quantifies the observed loss when reducing the Gaussian vector metric to the per-bit product metric~\fref{eq:productmetric}; in general, however, these two mismatched metrics need not be ordered and this difference can be negative.
Thus, these two quantities do not intrinsically decompose the information discarded by the detector's summary and its interface. The reason is that~$R_{\up{vec}}$ measures one particular metric, not the summary itself: a receiver that carefully post-processes the summary might achieve a rate above~$R_{\up{vec}}$ (cf.~\fref{rem:realize}).

Two properties of the VMD rate follow immediately from the above results.
First, by \fref{defi:gmi}, the VMD rate is achievable for every summary $\{\hat{\vec{s}}_\Reals,\hat{\mat{C}}_\Reals\}$, irrespective of how miscalibrated it might be. 
Second, by \fref{lem:divergence}(i), the VMD rate satisfies $R_{\up{vec}}\leq I(\vec{s};\vec{y})$ with equality for summaries whose Gaussian metric, raised to some power $\tau\geq0$, is proportional to the posterior.
Furthermore, the scaling parameter $\tau$ plays the same diagnostic role as in \fref{rem:sparam}.

\begin{rem}
\label{rem:scaling}
The scaling parameter~$\tau$ appears in all of our rate expressions and in the experiments of \fref{sec:results}.
In the GMI, $\tau$ is a parameter of the bound and we measure the rate at its maximizer $\tau^\star$; by \fref{lem:divergence}(i), this scaling minimizes the expected KL divergence from the true posterior to the normalized metric.
The same calibration issue affects a channel decoder because it trusts its inputs (cf.\ \fref{sec:decoder}); overconfident or underconfident soft outputs therefore degrade its performance.
Scaling the soft inputs by a constant $\tau$ before decoding is frequently used to mitigate this issue.\footnote{Corrections of mismatched soft outputs have been studied in~\cite{vandijk03,szczecinski12}.}
Thus, in our error-rate simulations, we scale the soft inputs prior to decoding.
\end{rem}

\begin{rem} \label{rem:whygmi}
The MI $I(\vec{s};\hat{\vec{s}})$ cannot distinguish between ``reasonable'' and ``unreasonable'' data detectors. 
Maximum-ratio combining (MRC), for example, which computes $\hat{\vec{s}}_{\up{MRC}}=\mat{H}\herm\vec{y}$, is arguably one of the crudest MIMO data detectors, but its output is a sufficient statistic for $\vec{s}$ (conditioned on~$\mat{H}$), so that $I(\vec{s};\hat{\vec{s}}_{\up{MRC}})=I(\vec{s};\vec{y})$; this is the maximum any data detector can achieve. 
The MI measures whether the information is still contained in the summary, whereas the vector GMI in~\fref{eq:vecgmi} asks what the Gaussian metric~\fref{eq:vecmetric} induced by that summary delivers to a decoder, which takes it at face value.
For MRC, a Gaussian model based on the error mean and covariance can differ from the posterior, so its vector GMI can be lower even though the estimate contains all information about the transmit vector.
\end{rem}

We next show that the metric~\fref{eq:vecmetric} is not arbitrary: for the ZF data detector, whose Gaussian error model is exact (cf.\ \fref{rem:zfexact}), the metric coincides with the exact posterior and achieves the bound $R_{\up{vec}}\leq I(\vec{s};\vec{y})$. 
We prove both statements, i.e., the posterior identity and the resulting rate, in \fref{app:natural}.

\begin{prop}[ZF Summary Is Information-Lossless] \label{prop:natural}
Assume that the Gram matrix $\mat{G}$ has full rank. Then, the posterior
in~\fref{eq:posterior} and the Gaussian decoding metric $q_{\up{vec}}$
in~\fref{eq:vecmetric} satisfy
\begin{align}  \label{eq:natkernel}
p(\vec{s}\,|\,\vec{y}) \propto q_{\up{vec}}\big(\vec{s}\,|\,\{\hat{\vec{s}}_{\up{ZF},\Reals},\hat{\mat{C}}_{\up{ZF},\Reals}\}\big),
\end{align}
for $\vec{s}\in\setO^{\MT}$, where $\{\hat{\vec{s}}_{\up{ZF}},\hat{\mat{C}}_{\up{ZF}}\}$ is the ZF summary in~\fref{eq:zf}. Consequently, the ZF summary achieves the maximum possible vector-level GMI~$R_{\up{vec}}=I(\vec{s};\vec{y})$ in \fref{eq:vecgmi}.
\end{prop}

In words, the exact posterior is a Gaussian metric of the form~\fref{eq:vecmetric}, because the ZF error model is exact (cf.\ \fref{rem:zfexact}).
This implies that a data detector does not need to be nonlinear to produce an information-lossless vector-level output in the sense of~\fref{eq:vecmetric}---it only needs to report the \emph{full} covariance matrix $\Nzero\mat{G}^{-1}$ together with the ZF estimate. 
In contrast, the LMMSE summary can be mismatched as its error contains residual interference from the discrete constellation, which may not be Gaussian (cf.\ \fref{rem:zfexact}). The following corollary, proved in \fref{app:tilt}, makes this explicit.

\begin{cor}[The Biased LMMSE Metric Is a Prior-Weighted Posterior] \label{cor:tilt}
Assume that the Gram matrix $\mat{G}$ has full rank and that $\Nzero>0$. 
Then, the Gaussian metric of the biased LMMSE summary in~\fref{eq:lmmse} satisfies
\begin{align}  \label{eq:tilt}
q_{\up{vec}}\big(\vec{s}\,|\,\{\hat{\vec{s}}_{\up{L},\Reals},\hat{\mat{C}}_{\up{L},\Reals}\}\big)\propto p(\vec{s}\,|\,\vec{y})\,\e^{-\|\vec{s}\|_2^2/\Es}.
\end{align}
Consequently, the LMMSE summary achieves $R_{\up{vec}}=I(\vec{s};\vec{y})$ if and only if the used constellation $\setO$ has constant modulus (e.g., PSK); otherwise, $R_{\up{vec}}<I(\vec{s};\vec{y})$.
\end{cor}

\begin{rem} \label{rem:realize}
The rate in~\fref{eq:vecgmi} is achievable by a decoder that scores codewords with the Gaussian vector metric~\fref{eq:vecmetric}, i.e., a decoder that operates on symbol vectors rather than bits.
BICM receivers do \emph{not} use such a decoder.
We use $R_{\up{vec}}$ solely as a benchmark for a data detector's summary, i.e., the rate that this summary supports with its own Gaussian vector metric~\fref{eq:vecmetric}, and not as the information contained in the summary.
A receiver that post-processes the summary with a better calibrated metric may exceed this rate, whereas no receiver can exceed $I(\vec{s};\vec{y})$.
\end{rem}

%%%
\subsection{The BICM Loss Is the Correlation of the Bits}
\label{sec:bicmloss}
We are now ready to identify where one-shot MIMO-BICM loses information.

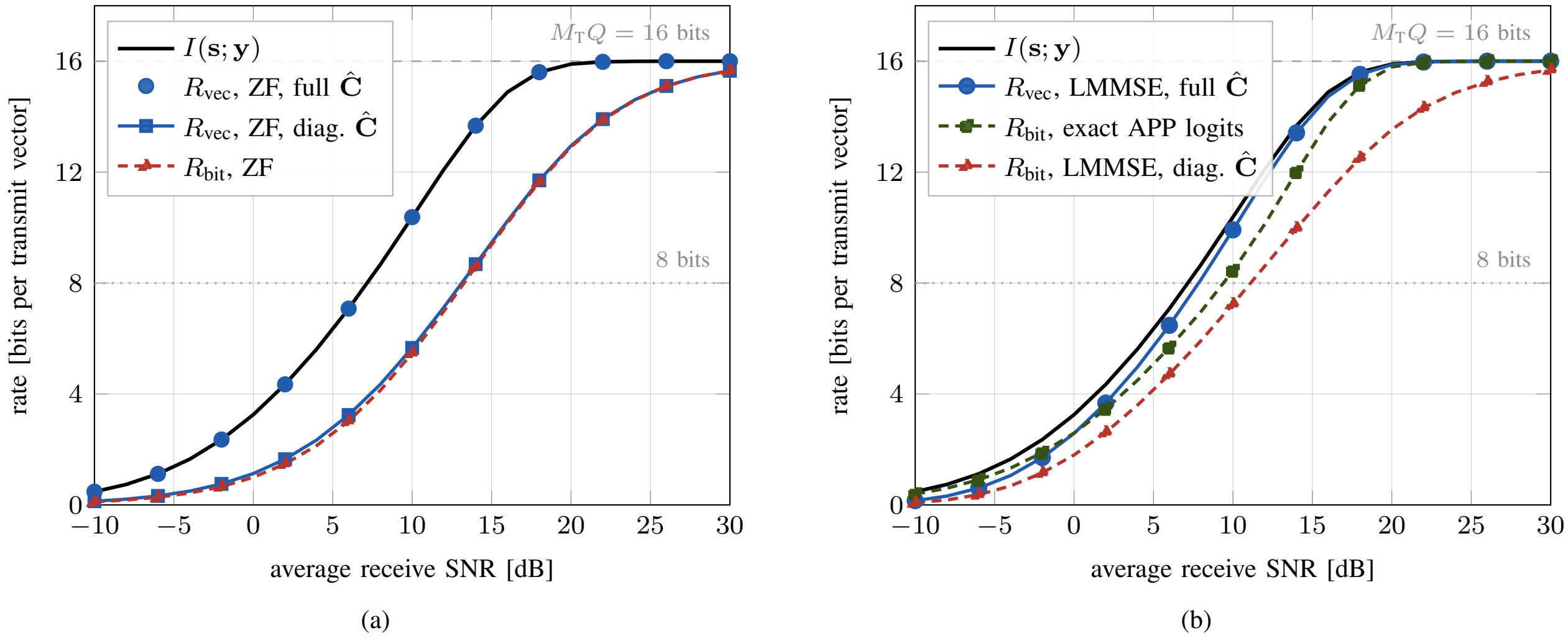
\begin{figure*}[tp]
\centering
\subfloat[\label{fig:zfladder}]{%
\begin{tikzpicture}
\begin{axis}[
  width=0.46\textwidth, height=0.38\textwidth,
  xlabel={average receive SNR [dB]},
  ylabel={rate [bits per transmit vector]},
  xmin=-10, xmax=30, ymin=0, ymax=18,
  xtick={-10,-5,...,30}, ytick={0,4,...,16},
  grid=both,
  grid style={line width=0.2pt, draw=black!12},
  major grid style={line width=0.2pt, draw=black!20},
  tick label style={font=\footnotesize},
  label style={font=\footnotesize},
  legend style={font=\footnotesize, at={(0.02,0.97)}, anchor=north west,
                draw=black!30, fill=white, fill opacity=0.9,
                text opacity=1, row sep=-1pt, inner sep=2.5pt},
  legend cell align={left},
]
% upper limit imposed by constellation MT*Q
\addplot[black!35, dashed, line width=0.5pt, forget plot]
  coordinates {(-10,16) (30,16)};
\node[font=\scriptsize, black!45, anchor=south east]
  at (axis cs:29.5,16.2) {$\MT Q=16$ bits};
% 8-bit reference limit (where we compute the operation point)
\addplot[black!35, dotted, line width=0.6pt, forget plot]
  coordinates {(-10,8) (30,8)};
\node[font=\scriptsize, black!45, anchor=south east]
  at (axis cs:29.5,8.15) {$8$ bits};
% upper bound: I(s;y)
\addplot[black, line width=1.1pt]
  table {\simdir/MI_4x4_16QAM_ZF_full_T1000_0.txt};
\addlegendentry{$I(\vec{s};\vec{y})$}
% Rvec for ZF summary with full covariance; this should be I(s;y) 
\addplot[only marks, mark=*, mark size=2.1pt, blue, line width=0.7pt,
         mark repeat=2]
  table {\simdir/Rvec_4x4_16QAM_ZF_full_T1000_0.txt};
\addlegendentry{$R_{\up{vec}}$, ZF, full $\hat{\mat{C}}$}
% Rvec, with diagonal ZF summary
\addplot[blue, line width=1.0pt, mark=square*, mark size=1.6pt, mark repeat=2]
  table {\simdir/Rvec_4x4_16QAM_ZF_diag_T1000_0.txt};
\addlegendentry{$R_{\up{vec}}$, ZF, diag.~$\hat{\mat{C}}$}
% Rbit with ZF summary and per-bit soft-outputs (equal for full and diagonal summaries; we take full here)
\addplot[red, densely dashed, line width=1.0pt, mark=triangle*, mark size=2pt,
         mark repeat=2]
  table {\simdir/Rbit_4x4_16QAM_ZF_full_T1000_0.txt};
\addlegendentry{$R_{\up{bit}}$, ZF \phantom{$\hat{\mat{C}}$}}
\end{axis}
\end{tikzpicture}}
\hfil
\subfloat[\label{fig:interface}]{%
\begin{tikzpicture}
\begin{axis}[
  width=0.46\textwidth, height=0.38\textwidth,
  xlabel={average receive SNR [dB]},
  ylabel={rate [bits per transmit vector]},
  xmin=-10, xmax=30, ymin=0, ymax=18,
  xtick={-10,-5,...,30}, ytick={0,4,...,16},
  grid=both,
  grid style={line width=0.2pt, draw=black!12},
  major grid style={line width=0.2pt, draw=black!20},
  tick label style={font=\footnotesize},
  label style={font=\footnotesize},
  legend style={font=\footnotesize, at={(0.02,0.97)}, anchor=north west,
                draw=black!30, fill=white, fill opacity=0.9,
                text opacity=1, row sep=-1pt, inner sep=2.5pt},
  legend cell align={left},
]
% upper limit imposed by constellation MT*Q
\addplot[black!35, dashed, line width=0.5pt, forget plot]
  coordinates {(-10,16) (30,16)};
\node[font=\scriptsize, black!45, anchor=south east]
  at (axis cs:29.5,16.2) {$\MT Q=16$ bits};
% 8-bit reference limit (where we compute the operation point)
\addplot[black!35, dotted, line width=0.6pt, forget plot]
  coordinates {(-10,8) (30,8)};
\node[font=\scriptsize, black!45, anchor=south east]
  at (axis cs:29.5,8.15) {$8$ bits};
% upper bound: I(s;y)
\addplot[black, line width=1.1pt]
  table {\simdir/MI_4x4_16QAM_ZF_full_T1000_0.txt};
\addlegendentry{$I(\vec{s};\vec{y})$}
% LMMSE with full covariance
\addplot[blue, line width=1.0pt, mark=*, mark size=2.1pt, mark repeat=2]
  table {\simdir/Rvec_4x4_16QAM_LMMSE_biased_full_T1000_0.txt};
\addlegendentry{$R_{\up{vec}}$, LMMSE, full $\hat{\mat{C}}$}
% soft-output APP with per-bit soft outputs (best possible per-bit detector)
\addplot[green_120, densely dashed, line width=1.0pt, mark=square*,
         mark size=1.6pt, mark repeat=2]
  table {\simdir/Rbit_4x4_16QAM_APP_full_T1000_0.txt};
\addlegendentry{$R_{\up{bit}}$, exact APP logits\phantom{$\hat{\mat{C}}$}\!\!\!}
%  LMMSE with per-bit soft outputs
\addplot[red, densely dashed, line width=1.0pt, mark=triangle*, mark size=2pt,
         mark repeat=2]
  table {\simdir/Rbit_4x4_16QAM_LMMSE_biased_full_T1000_0.txt};
\addlegendentry{$R_{\up{bit}}$, LMMSE, diag.~$\hat{\mat{C}}$}
\end{axis}
\end{tikzpicture}}
\caption{Achievable rates for a $4\times 4$ Rayleigh-fading MIMO system with 16-QAM.  
(a)~Fixed data detector with varied output format: We see that the output format determines the rate loss: the full ZF summary is information-lossless; reducing it to a diagonal covariance costs $6.0$\,dB SNR at $8$\,bits per transmit vector, and further reduction to per-bit soft outputs only $0.1$\,dB more.
(b)~Data detector and format are varied: We see that a better interface outperforms a better data detector, i.e., reporting the full LMMSE covariance gains $3.3$\,dB, whereas replacing LMMSE by the optimal APP data detector gains only $1.7$\,dB.}
\label{fig:ladders}
\end{figure*}

%%%
\subsubsection{Insight by Analysis}
We first identify, exactly, what a per-bit interface loses, even for an optimal data detector. 
We then show simulations to quantify the loss.
The proof of the following result is given in \fref{app:bicmloss}.

\begin{prop}[BICM Loss]
\label{prop:bicmloss}
With uniform i.i.d.\ coded bits and exact per-bit logits $\tilde{\lambda}_k=\lambda_k$ for $k=1,\ldots,N$, we have
\begin{align}
  I(\vec{s};\vec{y}) - R_{\up{bit}}
  = \sum_{k=1}^{N} H(x_k\,|\,\vec{y}) - H(\vec{x}\,|\,\vec{y}),
  \label{eq:bicmloss}
\end{align}
which is the conditional total correlation~\cite{watanabe60} of the coded bits of one
transmit vector given the receive vector.
\end{prop}

The expression in~\fref{eq:bicmloss} implies the following. With exact logits, the per-bit metrics are the true marginals: $q_k(x_k)=p(x_k\,|\,\vec{y})$.
Bit-metric decoding therefore scores every codeword with the product metric $\prod_{k=1}^{N}p(x_k\,|\,\vec{y})$, i.e., treats the $N$ bits of a transmit vector as \emph{independent}.
The right-hand side of~\fref{eq:bicmloss} is the average KL divergence between the true joint posterior and this product, i.e., $\Ex[\mathrm{KL}(p(\vec{x}\,|\,\vec{y})\,\|\,\textstyle\prod_{k=1}^{N}p(x_k\,|\,\vec{y}))]$, where the expectation is over the receive vector. This divergence is exactly the rate cost of assuming independence.
The per-bit product metric discards dependence among the bits, even if the data detector computes exact logits.

\begin{rem} \label{rem:classicalloss}
The identity in~\fref{eq:bicmloss} restates the classical BICM loss in terms of the total correlation~\cite{watanabe60}. 
The classical statement is the gap between the coded-modulation rate $I(\vec{s};\vec{y})$ and the sum $\sum_{k}I(x_k;\vec{y})$ of the rates of the parallel bit channels that the BICM interface induces. 
This sum is the BICM capacity of~\cite[Eq.~14]{caire98}, which is no larger than the coded-modulation capacity of~\cite[Eq.~10]{caire98}; this inequality is~\cite[Eq.~16]{caire98} and follows from the data-processing inequality.
Furthermore, \cite[Thm.~2 and Cor.~1]{martinez09} show that this sum is also the GMI of a bit-metric decoder fed with exact logits. Thus, the bit-level interface, rather than metric mismatch, causes the rate loss.
\end{rem}

%%%
\subsubsection{Quantification by Simulation}
We now quantify the loss identified by \fref{prop:bicmloss}. 
We simulate MI and GMI in a $\MT=\MR=4$ Rayleigh-fading MIMO system with 16-QAM and Gray labeling, where the entries of~$\mat{H}$ are i.i.d.\ with unit variance.
We measure all rates in bits per transmit vector. They are therefore upper bounded by $N=\MT Q=16$ bits, and we compute them using $1\,000$ Monte-Carlo trials.

In \fref{fig:zfladder}, we fix the data detector and vary only its output format.
From \fref{prop:natural}, we already know that the vector-level GMI of the ZF summary in~\fref{eq:zf} coincides with $I(\vec{s};\vec{y})$ at every SNR, so it does not lose information.
Reducing the summary to a diagonal covariance matrix costs $6.0$\,dB of SNR at $8$\,bits per transmit vector; a further reduction to per-bit soft outputs costs only an additional~$0.1$\,dB.
Put simply, with Gaussian summaries per~\fref{defi:eqdet}, discarding the off-diagonal entries of~$\hat{\mat{C}}$ accounts for almost the entire loss at the interface.

In \fref{fig:interface}, we vary the interface and the data detector. 
We start from the biased LMMSE data detector of \fref{sec:lmmse} with per-bit soft outputs; replacing it by the optimal APP data detector, i.e., the detector that computes the exact per-bit logits of~\fref{eq:exactlogit}, requires $1.7$\,dB less SNR for $8$ bits per transmit vector. 
Keeping the LMMSE data detector but reporting its full summary instead requires $3.3$\,dB less, and therefore outperforms the APP data detector by $1.6$\,dB. 
These results show that linearity and approximate per-bit logits are not the main limitations of a linear data detector. Instead, the main limitation is discarding the off-diagonal entries of the error covariance matrix.
In fact, the $3.3$\,dB gap between the per-bit and the full-summary interface is much larger than the $1.7$\,dB gap between the per-bit LMMSE and the optimal APP data detector.
Finally, the rates $I(\vec{s};\vec{y})$ and $R_{\up{bit}}$ with exact APP logits differ by the conditional total correlation in \fref{prop:bicmloss}. 
At $8$ bits per transmit vector, this rate loss corresponds to a $2.2$\,dB SNR gap in \fref{fig:interface}.

This loss can only be recovered if the channel decoder has access to inter-bit correlation within a transmit vector, which is discarded by the per-bit interface.
We therefore design a new soft-output format that satisfies the following four requirements:
(i)~it gives bit-metric decoders access to correlation information,
(ii)~it is simple to compute,
(iii)~it entails little storage overhead compared to conventional soft outputs, and
(iv)~it can be incorporated easily into existing BP-based channel decoders.

%%%%%%%%%%%%%%%%%%%%%%%%%%%%%%%%%%%%%%%%%%%%%%%%%%%%%%%%%%%
%% COIL-based data detector
%%%%%%%%%%%%%%%%%%%%%%%%%%%%%%%%%%%%%%%%%%%%%%%%%%%%%%%%%%%

\section{COIL-Based Data Detector}
\label{sec:coil}

COIL is a novel soft-output format in which the data detector passes per-bit soft outputs and pairwise \emph{coupling logits} to the channel decoder. The coupling logits convey correlation information between coded bits of different streams within a transmit vector---exactly the information whose removal dominates the loss of one-shot MIMO-BICM.
We first introduce the \emph{tree metric}, a codeword score that extends bit-metric decoding by one correction factor per selected bit pair.
We then show how equalization-based MIMO data detectors can compute these correction factors, how the relevant bit pairs can be selected, and how the result is represented as the COIL soft output.
We finally introduce the tree-metric GMI, which enables us to evaluate COIL directly at the rate level.

%%%
\subsection{The Tree Metric}
\label{sec:treemetric}

Our goal is to design the simplest refinement of bit-metric decoding that captures correlation between the coded bits of a transmit vector.
A bit-metric decoder does not model dependence among bits because its metric~\fref{eq:productmetric} factorizes them.
The minimal refinement therefore captures \emph{pairwise} correlation: for selected bit pairs \mbox{$(k,l)\in\setE$}, we multiply the product metric~\fref{eq:productmetric} by a \emph{correction factor}
\begin{align} \label{eq:correction}
c_{k,l}(x_k,x_l) \eqdef \frac{q_{k,l}(x_k,x_l)}{q_k(x_k)\,q_l(x_l)},
\end{align}
which compares a \emph{bit-pair metric} $q_{k,l}(x_k,x_l)$ that scores the two bits jointly with the per-bit metrics $q_k$ and $q_l$.
The correction factor is equal to one for all four sign combinations if and only if the bit-pair metric factorizes as $q_{k,l}(x_k,x_l)=q_k(x_k)\,q_l(x_l)$.
More generally, the two bits are independent under the normalized bit-pair metric exactly when the correction factor is a function of~$x_k$ times a function of~$x_l$.
Therefore, a correlated pair requires a correction factor that does not separate into per-bit factors.

We choose the set $\setE$ of bit pairs to form a \emph{spanning tree} (or a forest): a bit may be connected to several others, but the selected edges are not allowed to form a cycle. We make this choice for the following four reasons.
First, the best tree is known: Chow and Liu~\cite{chowliu68} showed that, among all tree-structured approximations of a given distribution, the best\footnote{Best in the sense of the smallest KL divergence to the distribution being approximated. Such an approximation keeps every pairwise dependence along the tree and discards everything else.} retains all per-variable marginals and the pairwise marginals along its edges. This tree is the maximum-weight spanning tree whose edge weights are the pairwise MIs.
Second, a tree metric can be normalized efficiently: the sum over the $2^N$ possible sign patterns of a transmit vector quickly becomes intractable for a general set $\setE$, but factorizes efficiently along a tree and requires only~$O(N)$ operations.
Third, trees are well suited to BP, which is exact on graphs without loops, whereas short cycles can negatively affect its convergence and accuracy. Correction factors placed along a tree form no loop on their own.\footnote{Every cycle must pass through the decoder's graph. Therefore, the interleaver affects the length of such cycles.}
Fourth, as we will show in~\fref{sec:receiver}, a BP-based decoder can process the correction factors using its usual update rules.

Concretely, we score the coded bits of one transmit vector using the following tree metric.
\begin{defi}[Tree Metric] \label{defi:treemetric}
Let $q_k$ be the per-bit metrics from~\fref{eq:productmetric}, let $q_{k,l}$ be strictly positive bit-pair metrics, let $c_{k,l}$ be the correction factors from \fref{eq:correction}, and let $\setE$ be the edge set of a spanning tree (or a forest) over the $N$ bits of one transmit vector. 
We define the \emph{tree metric} as follows:
\begin{align} \label{eq:treemetric}
q_{\up{tree}}(\vec{x}) = \prod_{k=1}^{N} q_k(x_k) \prod_{(k,l)\in\setE} c_{k,l}(x_k,x_l).
\end{align}
\end{defi}

A decoder scores an entire codeword with the product of~\fref{eq:treemetric} over the $K$ resource blocks, exactly as a bit-metric decoder scores it with the product of the per-bit metrics in~\fref{eq:productmetric}.

\begin{rem}
If every bit-pair metric factorizes into the two per-bit metrics, i.e., if $q_{k,l}(x_k,x_l)=q_k(x_k)\,q_l(x_l)$, then all correction factors~\fref{eq:correction} equal one and the tree metric~\fref{eq:treemetric} reduces to the product metric~\fref{eq:productmetric}. 
Thus, conventional bit-metric decoding is a special case of the tree metric in the absence of any correlation between the coded bits of a transmit vector.
\end{rem}

To turn~\fref{defi:treemetric} into the COIL soft-output format, we require three ingredients: (i)~compute the correction factors from the data detector's summary and reduce them to three parameters (see \fref{sec:coilcomp}); (ii)~score the candidate bit pairs and construct the spanning tree $\setE$ (see \fref{sec:treeconstr}); and (iii)~pass the resulting COIL soft output to the channel decoder (see \fref{sec:coildef}).

%%%
\subsection{Bit-Pair Metrics and Coupling Logits}
\label{sec:coilcomp}

We now show how a MIMO data detector in the sense of \fref{defi:eqdet} can compute the bit-pair metrics~$q_{k,l}$ of~\fref{eq:correction} from its summary $\{\hat{\vec{s}}_\Reals,\hat{\mat{C}}_\Reals\}$. We proceed in two steps.
\subsubsection{Two-Stream Gaussian Metric} 
For a bit pair $(k,l)$ from different streams $n(k)\neq n(l)$, extract the mean $\vec{m}_\Reals\in\Reals^4$ and covariance matrix~$\mat{\Sigma}_\Reals\in\Reals^{4\times4}$ of the two associated streams~$n(k)$ and $n(l)$ from the summary $\{\hat{\vec{s}}_\Reals,\hat{\mat{C}}_\Reals\}$, and form the two-stream Gaussian metric as in~\fref{eq:vecmetric}:
 \begin{align} \label{eq:pairkernel}
\phi_{k,l}(a,a') \eqdef  \e^{-\frac{1}{2}(\vec{v}_\Reals-\vec{m}_\Reals)\trans  \mat{\Sigma}_\Reals^{-1}(\vec{v}_\Reals-\vec{m}_\Reals)}.
 \end{align}
 Here, $\vec{v}=[a,a']\trans$ with $(a,a')\in\setO^2$, and $\vec{v}_\Reals\in\Reals^4$ is its real-valued decomposition.

\subsubsection{Bit-Pair Metrics} 
We marginalize\footnote{The sums in~\fref{eq:pairmarginal} can be approximated (cf.~\fref{sec:cheapweight}); using approximate scores limits the number of exact marginalizations to the selected tree edges.}~\fref{eq:pairkernel} to the bit pair using the label sets defined in~\fref{sec:model}:
\begin{align} \label{eq:pairmarginal}
q_{k,l}(x_k,x_l) = \sum_{a\in\setO_{b(k)}^{x_k}} \sum_{a'\in\setO_{b(l)}^{x_l}} \phi_{k,l}(a,a').
\end{align}
Computing each bit-pair metric therefore requires one marginalization over~$|\setO|^2$ terms, irrespective of $\MT$.
The number of bit pairs to be considered, and hence the total cost, grows with $\MT$ (cf.~\fref{sec:props}). 
For $\MT=2$, \fref{eq:pairmarginal} is exactly the bit-pair marginal of the vector metric~\fref{eq:vecmetric}; for $\MT>2$ it is an approximation of that marginal, which avoids a sum over $|\setO|^{\MT}$ terms.

\begin{rem}
We restrict the bit pairs of the two-stream Gaussian metric~\fref{eq:pairkernel} to different streams, as the interface loss is mostly due to the off-diagonal entries of~$\hat{\mat{C}}$ (cf.~\fref{fig:zfladder}), which characterize correlation among different streams. 
However, our construction also applies to arbitrary bit pairs, including pairs within one symbol; this may be relevant for other bit labelings (cf.~Footnote~\ref{fn:labeling}).
\end{rem}

We now represent the correction factors~\fref{eq:correction} in a compact format. 
The strictly positive bit-pair metric $q_{k,l}$ takes four values, one per sign combination $(x_k,x_l)\in\{\pm1\}^2$, and hence has four degrees of freedom---one of which is an irrelevant scale. 
The remaining three can be parametrized as an Ising model~\cite{ising25}, in which every spin (here: a coded bit) experiences a local \emph{field} that pulls it toward one of its two values, and every spin pair experiences a \emph{coupling} (here: their correlation) that rewards or penalizes agreement of the two spins.
We adopt this terminology and refer to the parameter that quantifies the coupling as the \emph{coupling logit}.
The following lemma makes this parametrization explicit; the proof is given in \fref{app:ising}.

\begin{lem}[Coupling-Logit Decomposition] \label{lem:ising}
For any strictly positive bit-pair metric $q_{k,l}$ and any strictly positive per-bit metrics $q_k$ and $q_l$, the correction factor~\fref{eq:correction} decomposes as
\begin{align}  \label{eq:isingfactor}
c_{k,l}(x_k,x_l) \propto \textstyle \e^{\frac{1}{2}\delta_{k}^{(k,l)} x_k + \frac{1}{2}\delta_{l}^{(k,l)} x_l + \frac{1}{2}\gamma_{k,l}\, x_k x_l},
\end{align}
with the \emph{coupling logit}
\begin{align} \label{eq:couplinglogit}
\gamma_{k,l} \eqdef \tfrac12 \ln \!\left(\frac{q_{k,l}(+1,+1)\,q_{k,l}(-1,-1)}{q_{k,l}(+1,-1)\,q_{k,l}(-1,+1)} \right)\!,
\end{align}
and the per-edge \emph{field corrections}
\begin{align}
\delta_{k}^{(k,l)} &\eqdef \tfrac12 \ln \!\left(\frac{q_{k,l}(+1,+1)\,q_{k,l}(+1,-1)}{q_{k,l}(-1,+1)\,q_{k,l}(-1,-1)}\right) - \tilde{\lambda}_k \label{eq:field} \\
\delta_{l}^{(k,l)} &\eqdef \tfrac12 \ln \!\left(\frac{q_{k,l}(+1,+1)\,q_{k,l}(-1,+1)}{q_{k,l}(+1,-1)\,q_{k,l}(-1,-1)}\right) - \tilde{\lambda}_l. \label{eq:fieldl}
\end{align}
Here, $\tilde{\lambda}_k$ and $\tilde{\lambda}_l$ are the conventional soft outputs that determine the per-bit metrics $q_k$ and $q_l$ in~\fref{eq:productmetric}. Equivalently, the bit-pair metric itself is the two-spin Ising model
\begin{align}
q_{k,l}(x_k,x_l)\propto \e^{h_{k}^{(k,l)}x_k+h_{l}^{(k,l)}x_l+\tfrac12\gamma_{k,l}x_kx_l},
\end{align}
with the \emph{Ising fields}
\begin{align}  \label{eq:isingfields}
h_{k}^{(k,l)} \eqdef \tfrac12\big(\tilde{\lambda}_k+\delta_{k}^{(k,l)}\big) \quad\text{and}\quad
h_{l}^{(k,l)} \eqdef \tfrac12\big(\tilde{\lambda}_l+\delta_{l}^{(k,l)}\big).
\end{align}
Since all of these quantities are invariant to a constant scaling of $q_{k,l}$, no extra normalization step is required.
\end{lem}

\fref{lem:ising} implies that a correction factor carries two types of information: the field corrections $\delta_k^{(k,l)}$ and $\delta_l^{(k,l)}$ describe the individual bits, whereas the coupling logit $\gamma_{k,l}$ describes the two bits jointly.
Only the coupling logits require a new interface in a MIMO-BICM receiver. A field correction acts on a single bit, so all field corrections collected by bit $k$ can be absorbed into its soft output. This approach yields the \emph{field-corrected soft output}:
\begin{align}  \label{eq:fieldcorrected}
\check{\lambda}_k \eqdef \tilde{\lambda}_k + \sum_{l:\{k,l\}\in\setE} \delta_{k}^{(k,l)}, \quad k=1,\dots,N.
\end{align}
In contrast, per-bit values cannot express the coupling logit: its factor $\e^{\gamma_{k,l}x_kx_l/2}$ rewards or penalizes only pairs of bit values.
The coupling logits are therefore the new information that a COIL-based data detector passes to the channel decoder. 
We will show in \fref{sec:receiver} how a BP-based LDPC channel decoder can incorporate this information efficiently, without changing the core algorithm.
The data detector therefore passes (i) the field-corrected soft outputs~$\check{\lambda}_k$ of~\fref{eq:fieldcorrected} and (ii) the coupling logits $\gamma_{k,l}$; by \fref{lem:ising}, this pair exactly represents the per-bit metrics and the correction factors---and hence the entire tree metric in~\fref{eq:treemetric}.

%%%
\subsection{Spanning Tree Construction}
\label{sec:treeconstr}

We now show how to select the bit pairs that form the spanning tree~$\setE$. We proceed in two steps: first, we first score every candidate bit pair by the amount of correlation; second, we construct the tree with the largest total score.

\subsubsection{Scoring the Candidate Pairs}
Chow and Liu~\cite{chowliu68} proposed a suitable score: the MI~$I_{k,l}$ between $x_k$ and~$x_l$ under the normalized bit-pair metric. Given the four values of the bit-pair metric, $I_{k,l}$ can be evaluated exactly; this, however, requires normalizing the bit-pair metric and evaluating several logarithms per candidate pair.
For weak coupling, the MI can be approximated using the Ising parameters of \fref{lem:ising}.
Concretely, expanding the MI of the bit pair to second order\footnote{For two weakly correlated binary random variables, the MI is $\mathrm{cov}(x_k,x_l)^2/(2\ln(2)\,\mathrm{var}(x_k)\mathrm{var}(x_l))$ up to higher-order terms; this is the local expansion of the KL divergence~\cite{kullback59}. In our case, we have $\mathrm{var}(x_k)=1-m_k^2$, and a first-order expansion with respect to the coupling logit yields $\mathrm{cov}(x_k,x_l)=\tfrac12\gamma_{k,l}(1-m_k^2)(1-m_l^2)$. Combining these expressions results in~\fref{eq:edgeweight}.}
in the coupling logit yields the following approximation:
\begin{align}  \label{eq:edgeweight}
I_{k,l} \approx \tilde{I}_{k,l} \eqdef \frac{\gamma_{k,l}^2}{8\ln(2)}\big(1-m_k^2\big)\big(1-m_l^2\big).
\end{align}
Here, the decoupled bit means of the Ising fields~\fref{eq:isingfields} are $m_k=\tanh(h_k^{(k,l)})$ and $m_l=\tanh(h_l^{(k,l)})$, i.e., the bit means at $\gamma_{k,l}=0$.
The score~\fref{eq:edgeweight} requires no normalization and corresponds to the squared coupling logit weighted by the uncertainty of the two decoupled bits.
Both the exact MI and the score~\fref{eq:edgeweight} still require the bit-pair metrics~\fref{eq:pairmarginal} of every candidate pair; we reduce this cost in \fref{sec:cheapweight}.

%%%
\subsubsection{Constructing the Tree}
\label{sec:treebuild} 
We evaluate the score $\tilde{I}_{k,l}$ for all candidate bit pairs, i.e., all cross-stream pairs of one transmit vector; there are at most $\MT(\MT-1)Q^2/2$ of them. We then construct the set~$\setE$ as the maximum-weight spanning tree over these scores by using Kruskal's algorithm~\cite{kruskal56}.

\begin{rem} \label{rem:treeselect}
Both steps can be done differently. Any score that ranks bit pairs by their correlation can replace $\tilde{I}_{k,l}$, and any tree-construction rule can
replace the maximum-weight spanning tree. We do not pursue such alternatives further. 
\end{rem}

%-----------------------------------------------------------
\subsection{Low-Complexity Spanning Tree Construction}
\label{sec:cheapweight}

The spanning-tree construction of \fref{sec:treeconstr} scores the $\MT(\MT-1)Q^2/2$ cross-stream candidate pairs using their bit-pair metrics~\fref{eq:pairmarginal}. 
Each score requires a marginalization over~$|\setO|^2$ terms.
Since the number of candidate pairs grows quadratically in the number of bits per transmit vector, these marginalizations dominate the complexity of COIL.
We therefore propose to use the max-log approximation, which replaces each sum in~\fref{eq:pairmarginal} by its largest term:
\begin{align}  \label{eq:maxlog}
\ln\big(q_{k,l}(x_k,x_l)\big) \approx \max_{a\in\setO_{b(k)}^{x_k}}\,\max_{a'\in\setO_{b(l)}^{x_l}} \ln\!\big( \phi_{k,l}(a,a')\big).
\end{align}
This approximation avoids the marginalizations.
Note that the number of candidate pairs itself stays quadratic in the number of bits per transmit vector; only the per-pair marginalization can be avoided. 
We write~$\tilde{\gamma}_{k,l}$ for the \emph{max-log coupling logit}, i.e., the coupling logit~\fref{eq:couplinglogit} evaluated with the approximation~\fref{eq:maxlog}.

Since $\ln(\phi_{k,l})$ is a negative quadratic form, each of the four maxima in~\fref{eq:maxlog} is attained by the symbol pair closest to the equalizer output under the corresponding bit hypothesis.
These four pairs satisfy $a\in\setO_{b(k)}^{x_k}$ and $a'\in\setO_{b(l)}^{x_l}$ for their respective bit hypotheses.
Closeness is measured in terms of distance induced by the matrix $\mat{\Sigma}_\Reals$ of~\fref{eq:pairkernel}. 
The coupling logit~\fref{eq:couplinglogit} combines the four values of $\ln(q_{k,l})$. 
Identifying these four maximizing pairs can still require a search over the constellation. 
We instead evaluate the metric at four pairs formed from independently selected nearest symbols and their bit-flipped versions.
The following \fref{lem:cheapgamma} gives the resulting coupling when these pairs attain the four maxima; otherwise, the expression is an additional approximation, evaluated in~\fref{tab:cheapcost}.

Before forming the precision matrix, we reorder the coordinates in~\fref{eq:pairkernel} as $[\Re(a),\Im(a),\Re(a'),\Im(a')]\trans$ and apply the same permutation to the mean and covariance. 
Let $\mat{P}$ denote the inverse covariance in this order, and partition it into four $2\times2$ blocks as
\begin{align} \label{eq:precblocks}
  \mat{P} = \begin{bmatrix} \mat{P}_{ii} & \mat{P}_{ij} \\
  \mat{P}_{ij}\trans & \mat{P}_{jj} \end{bmatrix}\!,
\end{align}
indexed by the two streams $i=n(k)$ and $j=n(l)$.
For stream~$i$ and bit $k$ with $n(k)=i$, we define the following quantities:
\emph{(i) the nearest symbol} 
\begin{align}
\hat{a}_i \eqdef \arg\min_{a\in\setO}|\hat{s}_i-a|, 
\end{align}
i.e., the nearest constellation point to the equalizer output at stream~$i$;
\emph{(ii) the nearest bit value} $\hat{x}_k\in\{\pm1\}$, i.e., the bit value of the $b(k)$th label associated with $\hat{a}_i$, so that $\hat{a}_i\in\setO_{b(k)}^{\hat{x}_k}$;
\emph{(iii) the flipped symbol} 
\begin{align}
\hat{a}_i^{(k)} \eqdef \arg\min_{a\in\setO_{b(k)}^{-\hat{x}_k}}|a-\hat{a}_i|,
\end{align}
i.e., the constellation point closest to $\hat{a}_i$ that carries the opposite value of bit~$k$; and
\emph{(iv) the difference} $\vec{\alpha}_k \eqdef (\hat{a}_i^{(k)}-\hat{a}_i)_\Reals\in\Reals^2$, i.e., the real-valued decomposition of the difference between the flipped symbol and the nearest symbol.
We define the additional four quantities $\hat{a}_j$, $\hat{x}_l$, $\hat{a}_j^{(l)}$, and $\vec{\beta}_l\eqdef(\hat{a}_j^{(l)}-\hat{a}_j)_\Reals$ analogously for stream $j$ and bit $l$.
The following lemma is proved in \fref{app:cheapgamma}.

\begin{lem}[Max-Log Coupling Logits] \label{lem:cheapgamma}
Let the four maxima in~\fref{eq:maxlog}, one per hypothesis $(x_k,x_l)$, be achieved at the four symbol pairs $(\hat{a}_i,\hat{a}_j)$, $(\hat{a}_i^{(k)},\hat{a}_j)$, $(\hat{a}_i,\hat{a}_j^{(l)})$, and $(\hat{a}_i^{(k)},\hat{a}_j^{(l)})$, i.e., at the nearest symbol pair and its flipped versions. 
Then, the max-log coupling logit is given by
\begin{align} \label{eq:cheapgamma}
\tilde{\gamma}_{k,l} = -\tfrac12\,\hat{x}_k\,\hat{x}_l\, \vec{\alpha}_k\trans \mat{P}_{ij}\, \vec{\beta}_l.
\end{align}
\end{lem}

The max-log coupling logit in~\fref{eq:cheapgamma} has two properties that matter in practice.
First, for QAM constellations with independently Gray-labeled in-phase and quadrature components, the bilinear form~$\vec{\alpha}_k\trans \mat{P}_{ij} \vec{\beta}_l$ uses only one entry of $\mat{P}_{ij}$ (see \fref{cor:graycheap} below). This property simplifies the computation.
Second, computing~\fref{eq:cheapgamma} requires only the equalizer's summary.

\begin{cor}[Separable Gray Labeling] \label{cor:graycheap}
For a QAM constellation with a separable Gray labeling, every bit is a function of either the in-phase or quadrature component. 
The difference then has a single nonzero entry: $\vec{\alpha}_k=d_k\vec{e}_{u(k)}$, where $u(k)\in\{1,2\}$ is the in-phase or quadrature axis of bit~$k$, $\vec{e}_1,\vec{e}_2$ are the coordinate vectors, and $d_k\in\Reals$ is the difference between the flipped symbol and the nearest symbol along that axis. 
Under the attainment condition of~\fref{lem:cheapgamma}, the max-log coupling logit corresponds to
\begin{align}\label{eq:graycheap}
\tilde{\gamma}_{k,l} = -\tfrac12\,\hat{x}_k\hat{x}_l\, d_k d_l \, [\mat{P}_{ij}]_{u(k),u(l)},
\end{align}
i.e., each max-log coupling logit is a single entry of the cross-stream precision block, scaled by the two differences and the appropriate sign.
\end{cor}

We note that $d_k$ is signed: for example, a sign bit whose nearest symbol has a positive coordinate flips to a negative coordinate, so $d_k<0$. 
The sign matters only when $\tilde{\gamma}_{k,l}$ is used as a value in the decoding metric; the tree construction uses only $\tilde{\gamma}_{k,l}^2$ in~\fref{eq:edgeweight} and is therefore unaffected.

\begin{rem} \label{rem:covfree}
The covariance matrix~$\mat{\Sigma}_\Reals$ can be extracted directly from an equalizer's full error covariance. Computing the matrix~$\mat{P}=\mat{\Sigma}_\Reals^{-1}$ requires one real-valued $4\times4$ matrix inversion per stream pair, after which the precision matrix can be reused for all $Q^2$ candidate bit pairs between those two streams.
An implementation that computes only the diagonal entries of the covariance matrix, as, e.g., in~\cite{haene06,studer11}, must additionally compute the required inter-stream blocks.
Moreover, by \fref{lem:cheapgamma}, the max-log coupling logit~$\tilde{\gamma}_{k,l}$ depends on the received signal only through the nearest symbols.
\end{rem}

\begin{rem}
\label{rem:selectcheap}
Our simulations show that the following approach achieves a favorable performance/complexity trade-off.
First, construct the spanning tree with the approximation in~\fref{eq:graycheap}. 
Then, compute the $N-1$ coupling logits of the tree exactly via~\fref{eq:pairmarginal} and \fref{eq:couplinglogit}. Candidate-pair scoring remains to be quadratic in the number of bits, but the complex marginalization step over~$|\setO|^2$ symbol pairs is performed only for the $N-1$ selected edges rather than for every candidate pair; see \fref{sec:results-cost}.

\end{rem}

%%%
\subsection{The COIL Soft Output}
\label{sec:coildef}
We are now ready to define the COIL soft-output format which the MIMO data detector passes to the channel decoder.

\begin{defi}[COIL Soft Output] \label{defi:coil}
The COIL soft output of an equalization-based MIMO data detector consists of
\begin{enumerate}
\item the bit-pair set $\setE$ of \fref{sec:treebuild},
\item the field-corrected soft outputs $\check{\lambda}_k$ of~\fref{eq:fieldcorrected} for $k=1,\dots,N$, and
\item the coupling logits $\gamma_{k,l}$ of~\fref{eq:couplinglogit} for $(k,l)\in\setE$,
\end{enumerate}
computed per transmit vector from the summary $\{\hat{\vec{s}}_\Reals,\hat{\mat{C}}_\Reals\}$.
\end{defi}

Specifically, a COIL-capable data detector replaces the $N$ conventional soft outputs by $N$ field-corrected soft outputs and adds at most $N-1$ coupling logits together with the bit-pair set~$\setE$ that represents the spanning tree.

\begin{rem}
COIL is detector-agnostic: any data detector in the sense of \fref{defi:eqdet} can compute these soft outputs. Detectors outside this class may also be used to compute coupling logits.
\end{rem}

%%%
\subsection{Tree-Metric GMI Evaluation of COIL}
\label{sec:tmd}

The bit-level GMI of \fref{sec:bitgmi} evaluates per-bit soft outputs directly without a code. Analogously, we can evaluate the GMI of the proposed COIL soft output directly. The tree metric~\fref{eq:treemetric} is a codeword score and therefore induces a mismatched-decoding GMI, as for~$R_{\up{bit}}$ in~\fref{eq:bitgmi} and $R_{\up{vec}}$ in~\fref{eq:vecgmi}.

\begin{defi}[Tree-Metric GMI]
\label{defi:tmd}
Let $q_{\up{tree}}(\vec{x})$ denote the tree metric~\fref{eq:treemetric} of one transmit
vector and let $R_{\up{tree}}\eqdef R(q_{\up{tree}})$ be the tree-metric decoding (TMD) rate. Then, we have
\begin{align}  \label{eq:tmdgmi}
R_{\up{tree}} = N + \max_{\tau\geq0}\>\Ex\!\left[\log_2 \! \left( \frac{q_{\up{tree}}(\vec{x})^\tau}{\sum_{\vec{x}'\in\{\pm1\}^{N}} q_{\up{tree}}(\vec{x}')^\tau} \right) \right]\!.
\end{align}
\end{defi}

The denominator of~\fref{eq:tmdgmi} sums the tree metric over all $2^N$ sign patterns. 
Since the metric factorizes along a tree, this sum can be computed efficiently---this is one of our reasons for choosing a tree in \fref{sec:treemetric}. Achievability follows directly from \fref{defi:gmi} with input alphabet~$\{\pm1\}^{N}$.

%%%
\subsection{What Can a Tree Interface Recover?}
\label{sec:oracletree}

We now answer two fundamental questions:
\emph{how much rate can a tree interface recover and which trees should we construct?}
To answer both questions, we assume for a moment that the bit-pair metrics are the exact posterior pair marginals. We call the resulting metric the ``oracle tree metric'' because computing the exact posterior pair marginals requires marginalizing~\fref{eq:posterior} over all $|\setO|^{\MT}$ transmit vectors, which is prohibitively complex. Practical detectors therefore approximate these marginals.
\fref{prop:oracletree} below answers the first question and \fref{cor:mwst} the second; both proofs follow from \fref{lem:divergence} and are given in \fref{app:oracletree} and \fref{app:mwst}, respectively.

\begin{prop}[Oracle Tree Rate]
\label{prop:oracletree}
Fix a spanning tree (or forest) $\setE$ over the $N$ bits and build the tree metric of \fref{defi:treemetric} from the exact posterior marginals, i.e., $q_k(x_k)=\Pr[x_k\,|\,\vec{y}]$ and $q_{k,l}(x_k,x_l)=\Pr[x_k,x_l\,|\,\vec{y}]$.
Then, $\tau=1$ is optimal and
\begin{align}  \label{eq:oracletree}
R_{\up{tree}}  = R_{\up{bit}} + \sum_{(k,l)\in\setE} I(x_k;x_l\,|\,\vec{y}),
\end{align}
where $R_{\up{bit}}$ is the BMD rate for exact logits in~\fref{eq:bitgmi}.
Furthermore, we have $I(x_k;x_l\,|\,\vec{y})=\Ex[I_{\vec{y}}(x_k;x_l)]$, where $I_{\vec{y}}(x_k;x_l)$ is the MI between the two bits under their joint posterior for a given~$\vec{y}$ and the expectation is over the receive vector~$\vec{y}$.

If, instead, the trees are constructed per receive vector, i.e., $\setE=\setE(\vec{y})$, then the same statement holds with the instantaneous quantities inside the expectation:
\begin{align} \label{eq:oracletreey}
R_{\up{tree}} = R_{\up{bit}} + \Ex\!\Bigg[\sum_{(k,l)\in\setE(\vec{y})} I_{\vec{y}}(x_k;x_l)\Bigg]\!.
\end{align}
\end{prop}

In words, the oracle tree metric recovers the MI of the bit pairs connected by its edge set $\setE$---but nothing else.
\fref{prop:bicmloss} demonstrates that the per-bit interface loses the total correlation of the coded bits. Therefore, the two results together  identify the part of this loss that can be recovered by a tree metric, i.e., the pairwise MI collected by its edge set, and the part that it cannot.
The following corollary reveals three consequences of \fref{prop:oracletree}; the proof is given in \fref{app:mwst}.

\begin{cor}[Rate-Optimal Trees]
\label{cor:mwst}
Consider oracle tree metrics as in \fref{prop:oracletree}, and let $\setE$ be any spanning tree or forest over the $N$ bits. 
Then, 
\begin{enumerate}
\item a single tree $\setE$ used for all receive vectors maximizes~\fref{eq:oracletree} if it is a maximum-weight spanning tree for the averaged weights $I(x_k;x_l\,|\,\vec{y})$;
\item a tree that is constructed per receive vector maximizes~\fref{eq:oracletreey} if, for all $\vec{y}$ (except for a set of probability zero), $\setE(\vec{y})$
is a maximum-weight spanning tree for the instantaneous weights $I_{\vec{y}}(x_k;x_l)$; and
\item in both cases, the expected weight of the tree satisfies
\begin{align} \label{eq:treeceiling}
\Ex\!\Bigg[\sum_{(k,l)\in\setE(\vec{y})}\!\!\! I_{\vec{y}}(x_k;x_l)\Bigg]\leq I(\vec{s};\vec{y})-R_{\up{bit}}. 
\end{align}
For a single tree that is used for all receive vectors, the edge set no longer depends on~$\vec{y}$, and the left-hand side is simply $\sum_{(k,l)\in\setE}I(x_k;x_l\,|\,\vec{y})$.
Equality in \fref{eq:treeceiling} holds if and only if the posterior is tree-structured on the constructed tree, i.e., if and only if the posterior itself corresponds to the tree metric~\fref{eq:treemetric}, so that the coded bits have no dependence beyond the pairs connected by~$\setE$.
\end{enumerate}
\end{cor}

\begin{figure}[tp]
\centering
\begin{tikzpicture}
\begin{axis}[
  width=0.46\textwidth, height=0.38\textwidth,
  xlabel={average receive SNR [dB]},
  ylabel={rate [bits per transmit vector]},
  xmin=-10, xmax=30, ymin=0, ymax=18,
  xtick={-10,-5,...,30}, ytick={0,4,...,16},
  grid=both,
  grid style={line width=0.2pt, draw=black!12},
  major grid style={line width=0.2pt, draw=black!20},
  tick label style={font=\footnotesize},
  label style={font=\footnotesize},
  legend style={font=\footnotesize, at={(0.02,0.97)}, anchor=north west,
                draw=black!30, fill=white, fill opacity=0.9,
                text opacity=1, row sep=-1pt, inner sep=2.5pt},
  legend cell align={left},
]
% ---- constellation-constrained ceiling MT*Q
\addplot[black!35, dashed, line width=0.5pt, forget plot]
  coordinates {(-10,16) (30,16)};
\node[font=\scriptsize, black!45, anchor=south east]
  at (axis cs:29.5,16.2) {$\MT Q=16$ bits};
% ---- the rate at which the dB readings in the text are taken
\addplot[black!35, dotted, line width=0.6pt, forget plot]
  coordinates {(-10,8) (30,8)};
\node[font=\scriptsize, black!45, anchor=south east]
  at (axis cs:29.5,8.15) {$8$ bits};

% ---- I(s;y): benchmark
\addplot[black, line width=1.1pt]
  table {\simdir/MI_4x4_16QAM_ZF_full_T1000_0.txt};
\addlegendentry{$I(\vec{s};\vec{y})$}

% ---- same detector, full covariance reported at the vector level
\addplot[blue, line width=1.0pt, mark=*, mark size=2.1pt, mark repeat=2]
  table {\simdir/Rvec_4x4_16QAM_LMMSE_biased_full_T1000_0.txt};
\addlegendentry{$R_{\up{vec}}$, LMMSE, full $\hat{\mat{C}}$}

% ---- COIL: same detector, tree metric
\addplot[bronze, line width=1.0pt, mark=diamond*, mark size=2.1pt,
         mark repeat=2]
  table {\simdir/Rtree_4x4_16QAM_LMMSE_biased_full_T1000_0.txt};
\addlegendentry{$R_{\up{tree}}$, LMMSE, COIL \phantom{$\hat{\mat{C}}$}\!\!\!}

% ---- best possible detector, conventional per-bit interface
\addplot[green_120, densely dashed, line width=1.0pt, mark=square*,
         mark size=1.6pt, mark repeat=2]
  table {\simdir/Rbit_4x4_16QAM_APP_full_T1000_0.txt};
\addlegendentry{$R_{\up{bit}}$, exact APP logits\phantom{$\hat{\mat{C}}$}\!\!\!}

% ---- today's receiver: LMMSE with per-bit soft outputs
\addplot[red, densely dashed, line width=1.0pt, mark=triangle*, mark size=2pt,
         mark repeat=2]
  table {\simdir/Rbit_4x4_16QAM_LMMSE_biased_full_T1000_0.txt};
\addlegendentry{$R_{\up{bit}}$, LMMSE, diag.~$\hat{\mat{C}}$}
\end{axis}
\end{tikzpicture}
\caption{Achievable rates for a $4\times4$ Rayleigh-fading MIMO system with 16-QAM. At $8$ bits per transmit vector, COIL extracted from the
biased LMMSE summary requires $2.0$\,dB less SNR than the same
detector's per-bit interface, and $0.3$\,dB less than a per-bit interface fed
by a detector that computes exact APP logits (and whose complexity is prohibitive in practice).}
\label{fig:treeladders}
\end{figure}
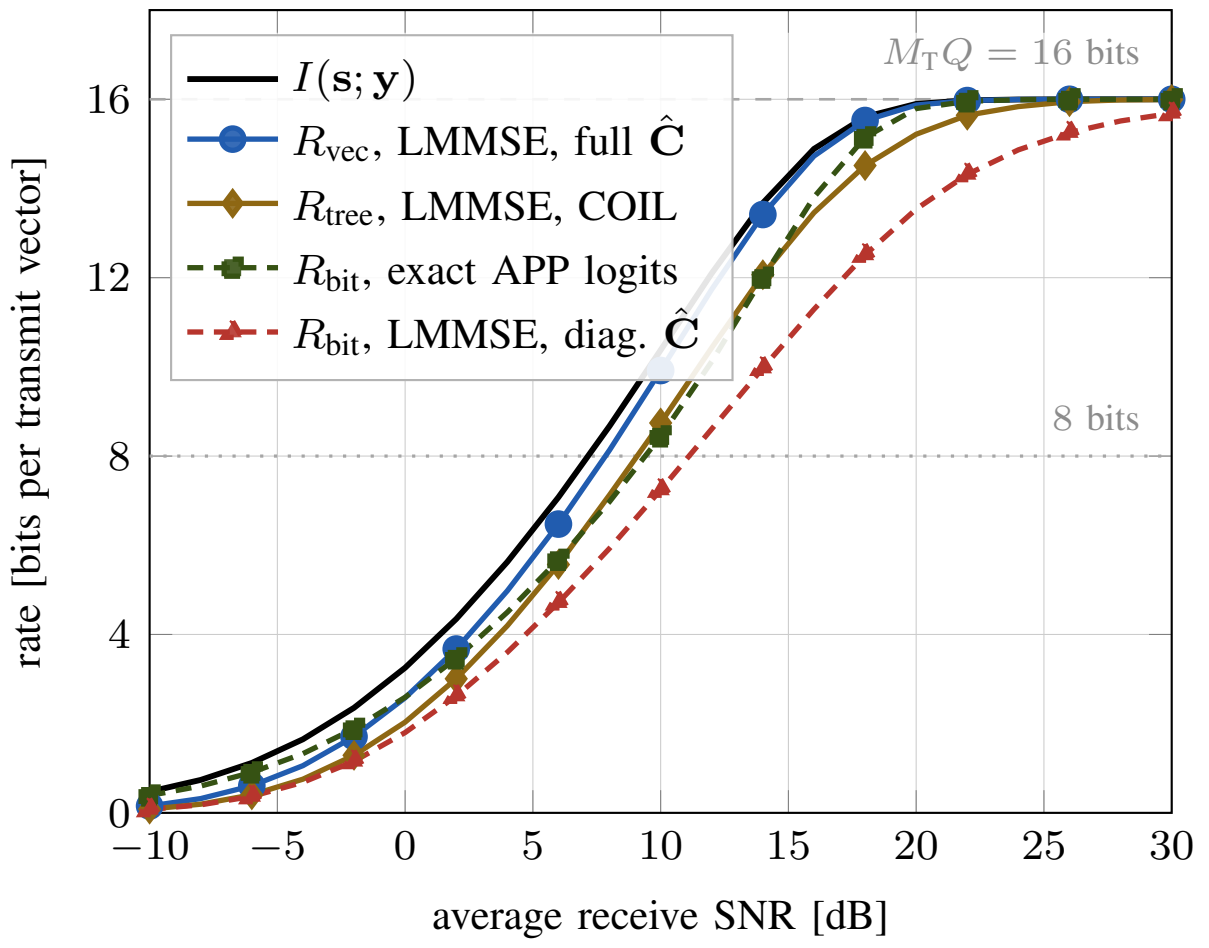

We emphasize that statement~(ii) is particularly relevant as COIL constructs a tree for every transmit vector.
Chow and Liu~\cite{chowliu68} motivate the maximum-weight spanning tree as the tree-structured PMF that is closest to the posterior; our corollary gives the
same tree a practical meaning, i.e., it is also the tree with the largest achievable rate.
For exact posterior marginals and MI weights, selecting a tree for each receive vector cannot give a lower rate than using a fixed tree. 
The gain is strict if no fixed tree maximizes the instantaneous weight sum for almost every receive vector. COIL uses the approximate score~\fref{eq:edgeweight}, so this guarantee might not hold for its selected trees.
Furthermore, statement~(iii) bounds what any tree interface can achieve: at most the entire conditional total correlation, and this is achieved only if the posterior happens to be tree-structured. 
Therefore, in \fref{fig:treeladders}, the remaining gap between~$R_{\up{tree}}$ and~$I(\vec{s};\vec{y})$ consists of two parts: the dependence that no tree can represent and the cost of the approximations used to generate COIL soft outputs.

\begin{rem} \label{rem:oracle}
\fref{prop:oracletree} is a best-case statement, and COIL can only approximate it as the bit-pair metrics are computed from the two-stream Gaussian summary of \fref{sec:coilcomp} instead of the exact posterior and the tree is constructed from the approximate score in~\fref{eq:edgeweight}.
Consequently, $R_{\up{tree}}\geq R_{\up{bit}}$ is guaranteed for the oracle tree metric, but not in general, i.e., inaccurate bit-pair metrics or a poorly constructed tree can make the rate gap~$R_{\up{tree}}-R_{\up{bit}}$ small or even negative.
\end{rem}

\begin{rem}
COIL can also use a fixed tree, selected once from the averaged weights in \fref{cor:mwst}(i), while updating its soft outputs for each receive vector.
This option might be useful in systems in which coupling is quasi-static, such as in optical links or in MIMO microwave backhaul or wired links with crosstalk.
A fixed tree avoids repeated tree construction, although even a fixed channel can produce receive-dependent bit-pair weights.
\end{rem}

%%%

We now empirically assess the tree metric.
We use the same setup as in \fref{sec:bicmloss}, i.e., we simulate a $\MT=\MR=4$ MIMO system with 16-QAM and Gray labeling over i.i.d.\ Rayleigh fading, and we measure the TMD rate $R_{\up{tree}}$. 

\fref{fig:treeladders} shows that the biased LMMSE data detector that computes COIL soft outputs from its full summary requires $2.0$\,dB less SNR (at $8$ bits per transmit vector) than the same detector's per-bit interface.
Furthermore, this method stays within $1.9$\,dB to $2.5$\,dB of $I(\vec{s};\vec{y})$ over the rate range from $4$\,bits to $14$\,bits per transmit vector.
More importantly, the coupling logits enable the LMMSE data detector to approach the performance of the optimal APP data detector: over the same range, an LMMSE detector with COIL operates within~$0.8$\,dB of a per-bit interface fed by \emph{exact} APP logits; at intermediate rates, COIL can even outperform this exact-APP per-bit interface (e.g., by $0.3$\,dB at $8$ bits per transmit vector), while avoiding the APP data detector's exhaustive search.

%%%%%%%%%%%%%%%%%%%%%%%%%%%%%%%%%%%%%%%%%%%%%%%%%%%%%%%%%%%
%% The COIL receiver
%%%%%%%%%%%%%%%%%%%%%%%%%%%%%%%%%%%%%%%%%%%%%%%%%%%%%%%%%%%

\section{COIL-Based Channel Decoder}
\label{sec:receiver}

We now complete our description of the COIL receiver architecture depicted in \fref{fig:rxcoil} by presenting the COIL-based channel decoder, which processes the COIL soft output of \fref{defi:coil}.
We first show how a BP-based LDPC channel decoder can process this new soft-output format. We then present the COIL receiver architecture, analyze its complexity, and compare it with one-shot and IDD receivers.

%-----------------------------------------------------------
\subsection{LDPC Decoding with COIL}
\label{sec:ldpc}

As a practically relevant example, we focus on LDPC decoding via BP~\cite{gallager62,kschischang01,richardson01}. 
We note, however, that our construction applies to any decoder that performs BP on a factor graph, including turbo decoders~\cite{kschischang01,loeliger04} and polar codes~\cite{arikan09} with BP decoding~\cite{arikan08,hussami09}.

The factor (Tanner) graph of a conventional LDPC code connects $\Ncw$ variable nodes (coded bits) to check nodes (parity checks), and the input interface consists of one logit (or soft input) per variable node; see \fref{fig:tannerconv}.
The key observation is that each coupling logit of \fref{defi:coil} can be included in the factor graph as one additional parity check whose third input is fixed to~$\gamma_{k,l}$; see \fref{fig:tannercoil}.
We simultaneously feed the field-corrected soft outputs~$\check{\lambda}_k$ to the decoder's regular inputs. Therefore, the factor graph remains bipartite, and the decoding algorithm remains unchanged.
The following results make these key insights precise; both proofs are given in \fref{app:boxplus} and~\fref{app:absorb}.

\begin{lem}[Coupling Logits Are Parity Checks with Fixed Input] \label{lem:boxplus}
Consider the pure coupling factor $\e^{\gamma_{k,l}x_k x_l/2}$ of~\fref{eq:isingfactor} and let $\mu$ be the logit of the BP message that this factor receives from variable node~$x_k$. 
Then, the message that the factor sends to variable node $x_l$ corresponds to the following logit:
\begin{align}  \label{eq:boxplus}
\gamma_{k,l} \boxplusop \mu \eqdef 2\tanh^{-1}\!\big(\tanh(\gamma_{k,l}/2)\tanh(\mu/2)\big).
\end{align}
Here, $\boxplusop$ is the well-known boxplus operator~\cite{hagenauer96}.
Thus, a coupling factor behaves exactly like a degree-$3$ parity check that connects $x_k$, $x_l$, and one additional degree-$1$ variable node whose
channel logit is fixed to~$\gamma_{k,l}$.
\end{lem}

\fref{lem:boxplus} only takes care of the coupling factor of~\fref{eq:isingfactor}; the correction factors, however, also contain the per-edge field corrections $\delta_k^{(k,l)}$ and $\delta_l^{(k,l)}$.
The following result shows that these field corrections require no additional check nodes, as they can be absorbed into the decoder's regular soft inputs.

\begin{lem}[Field Corrections Absorb into Channel Logits]
\label{lem:absorb}
The field corrections can be directly included into the channel logits. 
Concretely, BP with the full correction factors~\fref{eq:isingfactor} gives the same bit beliefs as BP with the \emph{pure coupling factors} $\e^{\gamma_{k,l}x_k x_l/2}$ when the schedule and message initialization correspond and every channel input $\tilde{\lambda}_k$ is replaced by $\check{\lambda}_k$ from~\fref{eq:fieldcorrected}.
The logits of messages on a coupling edge differ by the field correction at that edge's bit endpoint.
\end{lem}

%%%
\subsection{COIL Receiver Architecture}
\label{sec:arch}

Combining \fref{lem:boxplus} and \fref{lem:absorb} leads to the receiver architecture shown in \fref{fig:rxcoil}. 
Decoding requires a \emph{preprocessing stage}, performed once per code, interleaver, modulation, and antenna configuration, and a \emph{runtime stage}, performed for each transmitted codeword.

\subsubsection{Preprocessing: Static Factor Graph Augmentation}
\begin{figure}[tp]
\centering
\subfloat[\label{fig:tannerconv}]{%
\begin{tikzpicture}[
  font=\footnotesize,
  var/.style={draw, circle, minimum size=4.8mm, inner sep=0pt, fill=white,
              line width=0.7pt, font=\scriptsize},
  chk/.style={draw, rectangle, minimum size=4.4mm, inner sep=0pt, fill=white,
              line width=0.7pt, font=\scriptsize},
  lbl/.style={font=\scriptsize, inner sep=1pt},
  >={Triangle[width=1.6mm,length=1.6mm]}
]
  \foreach \i in {1,...,6} { \node[var] (a\i) at (0.78*\i+0.85,0) {$=$}; }
  \node[chk] (ac1) at (2.41,1.05) {$+$};
  \node[chk] (ac2) at (3.71,1.05) {$+$};
  \node[chk] (ac3) at (5.01,1.05) {$+$};
  \begin{scope}[on background layer]
  \foreach \c/\v in {ac1/a1,ac1/a2,ac1/a4,ac2/a2,ac2/a3,ac2/a5,
                     ac3/a3,ac3/a5,ac3/a6} { \draw (\c) -- (\v); }
  \end{scope}
  \foreach \i in {1,...,6} { \draw[->] (0.78*\i+0.85,-0.75) -- (a\i); }
  \node[lbl, anchor=north] at (1.63,-0.79) {$\tilde{\lambda}_k$};
  \path (0.30,-1.10) rectangle (5.45,1.35);
\end{tikzpicture}}
\\[0.10cm]
\subfloat[\label{fig:tannercoil}]{%
\begin{tikzpicture}[
  font=\footnotesize,
  var/.style={draw, circle, minimum size=4.8mm, inner sep=0pt, fill=white,
              line width=0.7pt, font=\scriptsize},
  chk/.style={draw, rectangle, minimum size=4.4mm, inner sep=0pt, fill=white,
              line width=0.7pt, font=\scriptsize},
  cchk/.style={draw, rectangle, minimum size=4.4mm, inner sep=0pt,
               line width=0.7pt, font=\scriptsize, fill=blue!12},
  lbl/.style={font=\scriptsize, inner sep=1pt},
  >={Triangle[width=1.6mm,length=1.6mm]}
]
\foreach \i in {1,...,6} { \node[var] (b\i) at (0.78*\i+0.85,0) {$=$}; }
\node[chk] (bc1) at (2.41,1.05) {$+$};
\node[chk] (bc2) at (3.71,1.05) {$+$};
\node[chk] (bc3) at (5.01,1.05) {$+$};
\begin{scope}[on background layer]
\foreach \c/\v in {bc1/b1,bc1/b2,bc1/b4,bc2/b2,bc2/b3,bc2/b5,
                     bc3/b3,bc3/b5,bc3/b6} { \draw (\c) -- (\v); }
\end{scope}
\foreach \i in {1,...,6} { \draw[->] (0.78*\i+0.85,-0.75) -- (b\i); }
\node[lbl, anchor=north] at (1.63,-0.79) {$\check{\lambda}_k$};
\node[cchk] (g1) at (0.45,1.05) {$+$};
\node[cchk] (g2) at (1.10,1.05) {$+$};
\draw[->] (0.45,-0.75) -- (g1);
\draw[->] (1.10,-0.75) -- (g2);
\node[lbl, anchor=north] at (0.775,-0.79) {$\gamma_{k,l}$};
\begin{scope}[on background layer]
\draw (g1) -- (b1);   \draw (g1) -- (b3);
\draw (g2) -- (b2);   \draw (g2) -- (b5);
\end{scope}
\path (0.30,-1.10) rectangle (5.45,1.35);
\end{tikzpicture}}
\caption{Illustration of a factor graph of an LDPC code (a) before and (b) after COIL augmentation. 
COIL incorporates each coupling logit by adding one parity check (shaded squares \cchknodesym\,) connected to the two coupled bits and the fixed input $\gamma_{k,l}$. Similarly, the field-corrected soft output $\check{\lambda}_k$ enters an ordinary variable node (circles \varnodesym\,).
Since the factor graph remains bipartite, the BP-based decoding algorithm remains the same.}
\label{fig:tanner}
\end{figure}
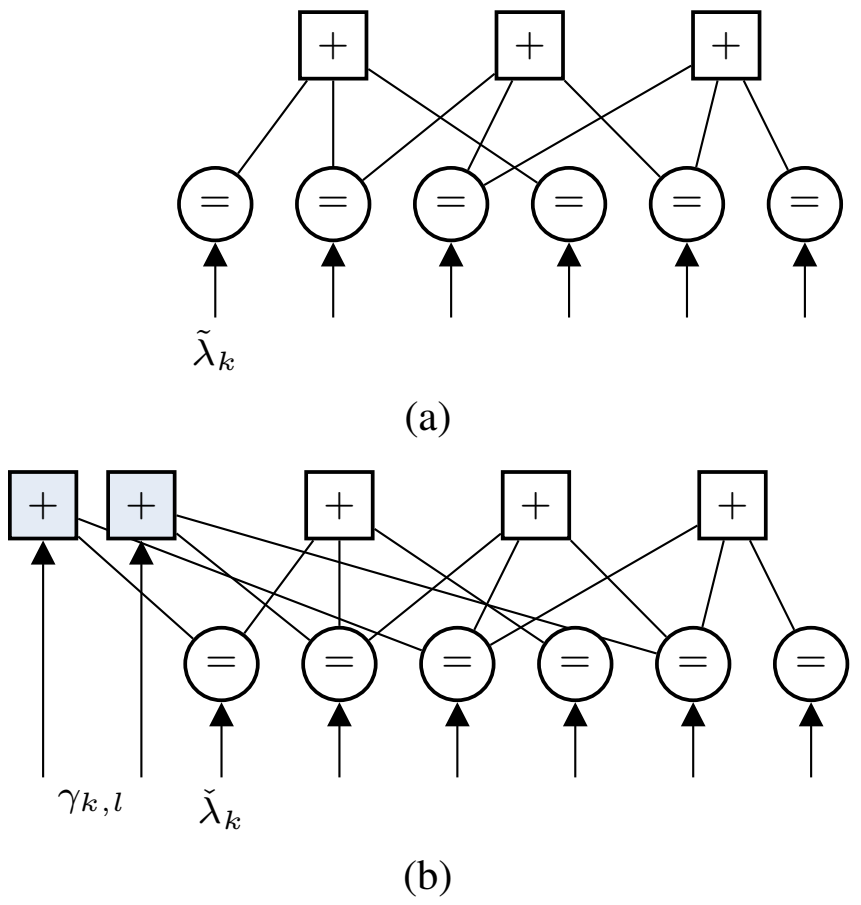

For every candidate cross-stream bit pair $(k,l)$ of a transmit vector, we add one new check node that connects $x_k$ and $x_l$ and that takes a fixed third
input. 
By \fref{lem:boxplus}, this is the degree-$3$ check with its degree-$1$ variable node absorbed into the fixed input, so the graph stays bipartite and no new variable nodes are created; see \fref{fig:tannercoil}.
The shaded part of the channel decoder in \fref{fig:rxcoil} indicates this augmentation.
This graph augmentation depends only on the code and the interleaver, but not on the channel realization, and is therefore \emph{static}.

\subsubsection{Runtime: COIL Detection and Decoding}
For each transmit vector, the receiver executes the following four steps:
\emph{(i) Equalization:} As for a conventional equalization-based data detector according to~\fref{defi:eqdet}, obtain the summary $\{\hat{\vec{s}}_\Reals,\hat{\mat{C}}_\Reals\}$, and compute the conventional soft outputs $\tilde{\lambda}_k$ from its diagonal as in~\fref{eq:symbolmetric} and~\fref{eq:logit}. 
\emph{(ii) COIL soft-output computation:} Compute the bit-pair metrics~\fref{eq:pairmarginal} of the candidate pairs from the off-diagonal entries of the summary; extract the coupling logits~$\gamma_{k,l}$ in~\fref{eq:couplinglogit} and the field corrections $\delta_k^{(k,l)}$ in~\fref{eq:field} and~\fref{eq:fieldl} using \fref{lem:ising}; score the candidate pairs with~\fref{eq:edgeweight} and construct the tree $\setE$ (cf.\ \fref{sec:treebuild}); and form the field-corrected soft outputs~$\check{\lambda}_k$ of~\fref{eq:fieldcorrected}. 
This is illustrated by the shaded part of the MIMO data detector in \fref{fig:rxcoil}.
\emph{(iii) Deinterleaving and soft-input computation:} Deinterleave the field-corrected soft outputs as usual and feed them to the ordinary variable nodes. Feed the coupling logits $\gamma_{k,l}$ to the fixed inputs of the corresponding coupling check nodes; set inactive (non-tree) pairs to~$\gamma_{k,l}=0$.
\emph{(iv) Decoding:} Run a standard BP algorithm on the augmented graph; this requires no change to the check- and variable-node updates or the message formats. Note that coupling checks with $\gamma_{k,l}=0$ can be skipped.

%%%

%-----------------------------------------------------------
\subsection{Complexity and Memory Requirements}
\label{sec:props}

COIL incurs overheads in terms of complexity and memory compared to conventional one-shot BICM receivers that exchange per-bit soft outputs only. 
We now discuss the added complexity at the data detector and at the channel decoder, and then compare COIL with IDD.

\subsubsection{Added Complexity in the Data Detector}
Compared to a conventional soft-output data detector, COIL performs two additional steps: (i) constructing the spanning tree $\setE$ and (ii) computing the coupling logits associated with the $N-1$ edges of the tree. 
Following \fref{rem:selectcheap}, tree construction should use the Gray-mapping approximation in~\fref{eq:graycheap} to compute each candidate coupling from one entry of the cross-stream precision block; the score also includes the uncertainty weights in~\fref{eq:edgeweight}.

Thus, all $\MT(\MT-1)Q^2/2$ candidate pairs can be scored without any marginalization, and Kruskal's algorithm can construct the maximum-weight spanning tree with these scores (cf.\ \fref{sec:treebuild}).
Also following \fref{rem:selectcheap}, the coupling logits associated with the $N-1$ edges of the tree should be computed exactly as in~\fref{eq:pairmarginal} and~\fref{eq:couplinglogit}; this requires marginalizing over~$|\setO|^2$ symbol pairs per edge, which dominates the added complexity.
We measure complexity by the number of metric computations per transmit vector. For COIL, these are evaluations of the symbol-pair kernel~\fref{eq:pairkernel}; for the exact one-shot APP detector, these are evaluations of one posterior term per transmit-vector hypothesis in~\fref{eq:exactlogit}. Under this measure, COIL requires~$(N-1)\,|\setO|^2$ metric computations, whereas the exhaustive-search APP detector requires $|\setO|^{\MT}$.
We reiterate that this count is for the low-complexity variant proposed in~\fref{rem:selectcheap}, i.e., max-log scoring of the candidate pairs followed by exact coupling logits on the $N-1$ tree edges.\footnote{Our error-rate simulations of \fref{sec:results} compute the candidate pairs exactly, which requires about $6\times$ more metric evaluations ($\MT(\MT-1)Q^2/2=96$ instead of $N-1=15$ marginalizations per transmit vector for $\MT=4$ and 16-QAM); \fref{tab:cheapcost} reports the loss when using approximate tree selection.}

As an example, consider a $\MT=\MR=4$ MIMO system with 16-QAM. In this case, COIL requires $3\,840$ metric evaluations compared to $65\,536$ for exact APP; for a $\MT=\MR=8$ system with 16-QAM, COIL requires $7\,936$ metric evaluations while exact APP requires $4.3\cdot10^9$.

With respect to memory consumption, COIL passes $2N-1$ soft values per transmit vector instead of $N$: the $N$ field-corrected soft outputs $\check{\lambda}_k$ plus the $N-1$ coupling logits $\gamma_{k,l}$ ($31$ instead of $16$ values for $\MT=\MR=4$ with 16-QAM). 
In addition, the decoder must know the bit pair associated with each coupling logit. This information can be encoded efficiently using $N-1$ indices for the tree edges.

\subsubsection{Added Complexity in the Channel Decoder}
Each active coupling check, i.e., each check with $\gamma_{k,l}\neq0$, has degree three with one fixed input; inactive checks are skipped. Only two of its three outgoing messages must therefore be computed, and each requires a single boxplus operation~\fref{eq:boxplus}.
With at most~$N-1$ active coupling checks per transmit vector, this increases the number of boxplus operations by at most $2(N-1)$ per transmit vector and BP iteration.

As an example, consider the rate-$1/2$ IEEE~802.11n~\cite{dot11n09} quasi-cyclic LDPC code of length $\Ncw=648$ that we also use in \fref{sec:results}. 
Assuming $3(d_{\up{c}}-2)$ boxplus operations per degree-$d_{\up{c}}$ check, the $324$ parity checks require $5\,184$ boxplus operations per iteration for both flooding and layered schedules. The coupling checks add at most $2(N-1)=30$ boxplus operations for each of the $\lceil\Ncw/N\rceil=41$ transmit vectors that carry one codeword. The resulting $1\,230$ additional boxplus operations correspond to a $24\%$ overhead.
Similar overheads apply for other standardized quasi-cyclic LDPC codes with similar check degrees, such as those of 5G~NR~\cite{nr38212}.

\subsubsection{Comparison with IDD}
\label{sec:idd}
IDD and COIL both exploit correlation between the bits within each transmit vector. IDD exploits it \emph{dynamically}: the channel decoder feeds back extrinsic or intrinsic (a posteriori) information, depending on the data-detector type~\cite{studer09}, and the data detector is re-run with these priors (cf.\ the dashed feedback path in \fref{fig:rxconv}). 
The complexity and latency of IDD therefore grow linearly in the number of outer iterations, and scheduling the processing of the two stages in hardware is difficult~\cite{studer09}.
COIL passes correlation information to the channel decoder once per transmit vector, and the decoder uses this information throughout decoding.
As a consequence, all iterations stay inside the channel decoder, avoiding repeated data-detector passes, feedback buffering, and scheduling between detector and decoder. 

%%%%%%%%%%%%%%%%%%%%%%%%%%%%%%%%%%%%%%%%%%%%%%%%%%%%%%%%%%%
%% Results
%%%%%%%%%%%%%%%%%%%%%%%%%%%%%%%%%%%%%%%%%%%%%%%%%%%%%%%%%%%

\section{Error-Rate Simulation Results}
\label{sec:results}

We now compare the full COIL receiver of \fref{sec:arch} with one-shot and IDD baselines in terms of frame error rate (FER).
We start by detailing the simulation scenario and then show FER results. We then investigate the underlying performance--complexity trade-off, study the impact of different system configurations, and finally quantify the cost of approximate coupling-logit computation.

%%%
\subsection{Simulation Scenario}
\label{sec:results-setup}

Our default simulation setup is as follows. 
We consider a Wi-Fi-like $\MT=\MR=4$ MIMO-BICM system with Gray-labeled 16-QAM and OFDM. 
For added realism, we simulate a four-tap frequency-selective channel with $K=41$ active subcarriers. The four time-domain taps are i.i.d.\ Rayleigh fading with a uniform power-delay profile.\footnote{The achievable rates of Secs.~\ref{sec:loss} and~\ref{sec:tmd} assume i.i.d.\ Rayleigh fading. This assumption remains valid as every subcarrier of the simulated system remains marginally i.i.d.\ Rayleigh. The short delay spread only results in correlation across subcarriers, which affects (i.e., reduces) frequency diversity but does not affect the per-subcarrier interface that is used by COIL.} We use the standardized rate-$1/2$ IEEE~802.11n quasi-cyclic LDPC code~\cite{dot11n09} with $\Ncw=648$ coded bits.\footnote{Since $\Ncw=648$ is not divisible by $N=16$, we pad the coded bits with eight zero bits, so that $K=41$ transmit vectors carry $656$ coded bits and the effective transmitted rate is $\Ninfo/(\Ncw+8)=324/656$.}
The coded bits pass through a random bit interleaver that is drawn once and reused for all Monte-Carlo trials.
The LDPC decoder uses a layered decoding schedule~\cite{hocevar04}, unless stated otherwise. 
We simulate the FER versus the average receive SNR, using at least $5\,000$ Monte-Carlo trials per SNR.

We compare four receivers with the same LDPC decoder:
\emph{(i) One-shot LMMSE:} we use the biased LMMSE data detector of \fref{sec:lmmse} with the conventional per-bit interface; this is a widely used detector in practice.
\emph{(ii) One-shot APP:} the optimal a posteriori probability data detector with the same per-bit interface; its soft outputs are the exact logits, so it maximizes the achievable rate~\fref{eq:bitgmi} among all one-shot detectors that use the standard per-bit product metric~\fref{eq:productmetric}. 
We use it as a benchmark in that sense, not as a FER bound for a finite-length receiver (cf.\ \fref{sec:limits}).
\emph{(iii) MMSE-based parallel interference cancellation (MMSE-PIC) with intrinsic feedback:} we use an IDD receiver that runs the MMSE-PIC data detector~\cite{studer11}, which builds upon soft interference cancellation followed by linear MMSE equalization~\cite{wang99,tuchler02mmse}, once per outer iteration, where $T$ is the number of outer iterations and $T=1$ denotes one-shot decoding. COIL aims to approach the performance gain of this practical IDD receiver without its high complexity and latency.
\emph{(iv) One-shot LMMSE with COIL:} we use the proposed receiver illustrated in \fref{fig:rxcoil}. This method builds on the data detector from~(i), uses the COIL soft outputs from \fref{defi:coil}, and employs the augmented factor graph of \fref{fig:tannercoil} in the LDPC decoder.

As mentioned in~\fref{rem:scaling}, we scale the soft inputs of each receiver by a constant $\tau$ before decoding, and we select this factor per configuration and per receiver.
For the default configuration with ten layered decoder iterations, COIL uses $\tau=0.85$ and the biased LMMSE receiver uses $\tau=0.90$.
For the unbiased LMMSE receiver, exact APP, and MMSE-PIC, we use $\tau=1$.

%%%
\begin{figure}[tp]
\centering
\begin{tikzpicture}
\begin{semilogyaxis}[
  scale only axis, width=0.81\columnwidth, height=0.72\columnwidth,
  xlabel={average receive SNR [dB]},
  ylabel={frame error rate (FER)},
  xmin=8, xmax=18, ymin=1e-3, ymax=1, 
  xtick={8,10,...,18},
  ytick={1e-4,1e-3,1e-2,1e-1,1},
  grid=both,
  grid style={line width=0.2pt, draw=black!12},
  major grid style={line width=0.2pt, draw=black!20},
  tick label style={font=\footnotesize},
  label style={font=\footnotesize},
  legend style={font=\scriptsize, at={(0.02,0.03)}, anchor=south west,
                draw=black!30, fill=white, fill opacity=0.9,
                text opacity=1, row sep=-1pt, inner sep=2.5pt},
  legend cell align={left},
]
% ---- target FER (the level at which the dB readings in the text are taken)
\addplot[black!35, dotted, line width=0.6pt, forget plot]
  coordinates {(8,1e-2) (18,1e-2)};
\node[font=\scriptsize, black!45, anchor=south east]
  at (axis cs:17.95,1.01e-2) {target FER};
\addplot[biaStyle]
  table {\simdirerr/FER_ERR_4x4_16QAM_4tap_LMMSE_bia_tau_LDPC_648bR12_lay_off_t090_L10_I1_1.txt};
\addlegendentry{biased LMMSE, per-bit}  
\addplot[unbStyle]
  table {\simdirerr/FER_ERR_4x4_16QAM_4tap_LMMSE_unb_LDPC_648bR12_lay_L10_I1_1.txt};
\addlegendentry{unbiased LMMSE, per-bit}  
\addplot[coilStyle]
  table {\simdirerr/FER_ERR_4x4_16QAM_4tap_COIL_LDPC_648bR12_lay_t085_n100000_L10_I1_1.txt};
\addlegendentry{COIL (this paper)}
\addplot[appStyle]
  table {\simdirerr/FER_ERR_4x4_16QAM_4tap_APP1_LDPC_648bR12_lay_n50000_L10_I1_1.txt};
\addlegendentry{exact APP, one shot}
\addplot[iddStyle]
  table {\simdirerr/FER_ERR_4x4_16QAM_4tap_MMSE_PIC_LDPC_648bR12_lay_n50000_L10_I4_1.txt};
\addlegendentry{IDD MMSE-PIC, $T\!=\!4$}
\end{semilogyaxis}
\end{tikzpicture}
\caption{FER vs. SNR for the considered receivers with ten layered LDPC decoder iterations.
At a target FER of $1$\%, one-shot COIL gains $1.3$\,dB over the best per-bit LMMSE receiver, operates within $0.7$\,dB of the exact one-shot APP detector, and operates within $0.8$\,dB of four-iteration IDD; note that IDD uses ten layered LDPC decoder iterations in each of its four outer iterations.}
\label{fig:fer}
\end{figure}
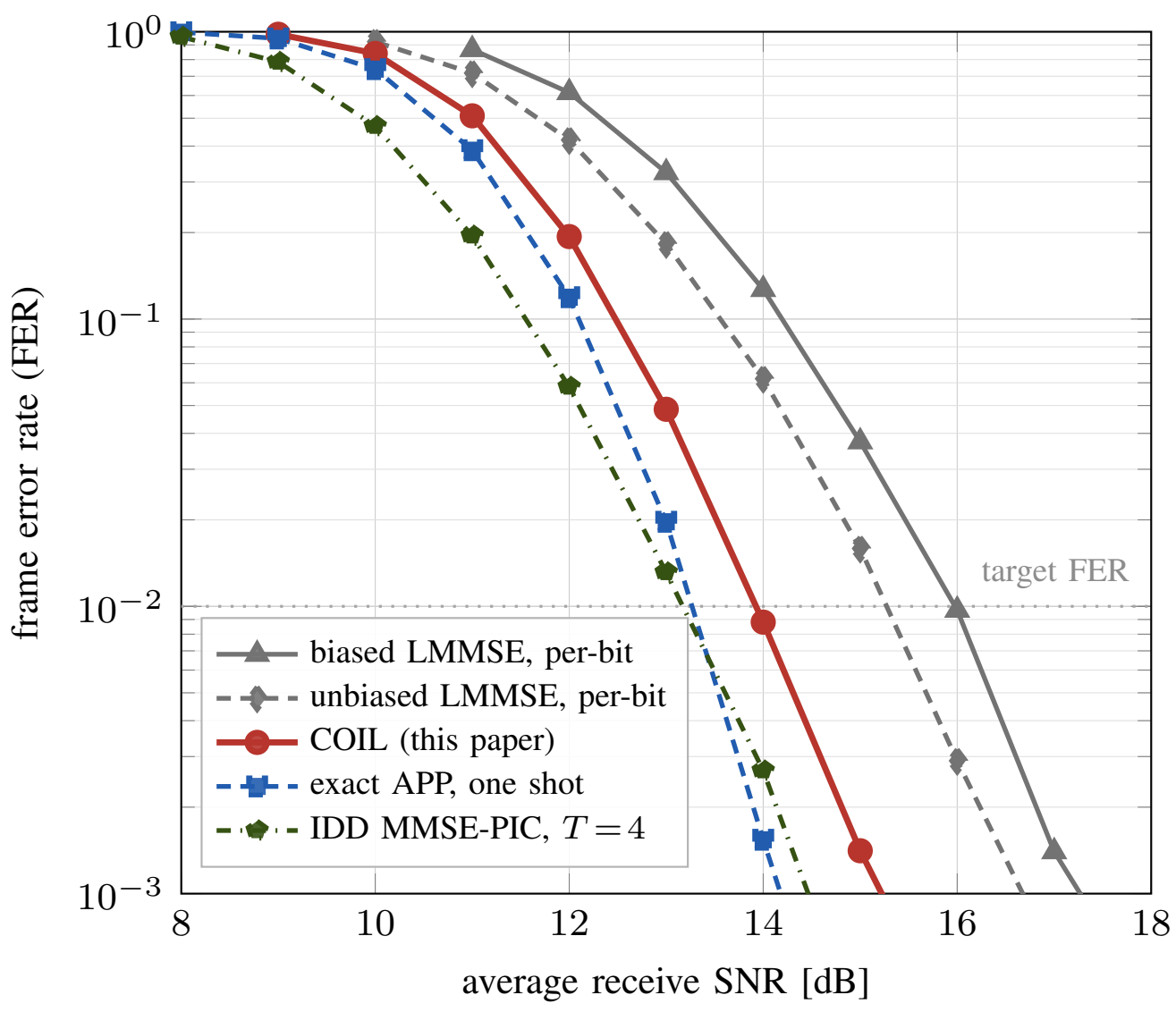

%%%
\subsection{FER Performance}
\label{sec:results-fer}

\fref{fig:fer} compares the FER of the four considered receivers, where the LDPC decoder performs ten layered iterations.
All three LMMSE-based curves use the same equalization stage. COIL and the biased per-bit receiver additionally share the same per-bit soft outputs, so these two curves differ only in their interface to the channel decoder.
We also compare two benchmarks: (i) the exact one-shot APP detector, which maximizes the achievable rate among all one-shot detectors that use the conventional per-bit interface; and (ii) MMSE-PIC with $T=4$ detector passes, where the channel decoder performs ten layered iterations in every outer iteration.

At a target FER of $1\%$, the COIL receiver outperforms the biased and unbiased per-bit LMMSE receivers by $2.0$\,dB and $1.3$\,dB, respectively. Furthermore, COIL operates within $0.7$\,dB of the exact one-shot APP detector and within $0.8$\,dB of IDD with four outer iterations.
We emphasize that COIL achieves this performance in a one-shot receiver.
The remaining gap to the exact one-shot APP detector decreases as the number of LDPC decoder iterations increases. As we show in \fref{sec:results-tradeoff}, the gap decreases to $0.10$\,dB with $40$ layered iterations.

%%%
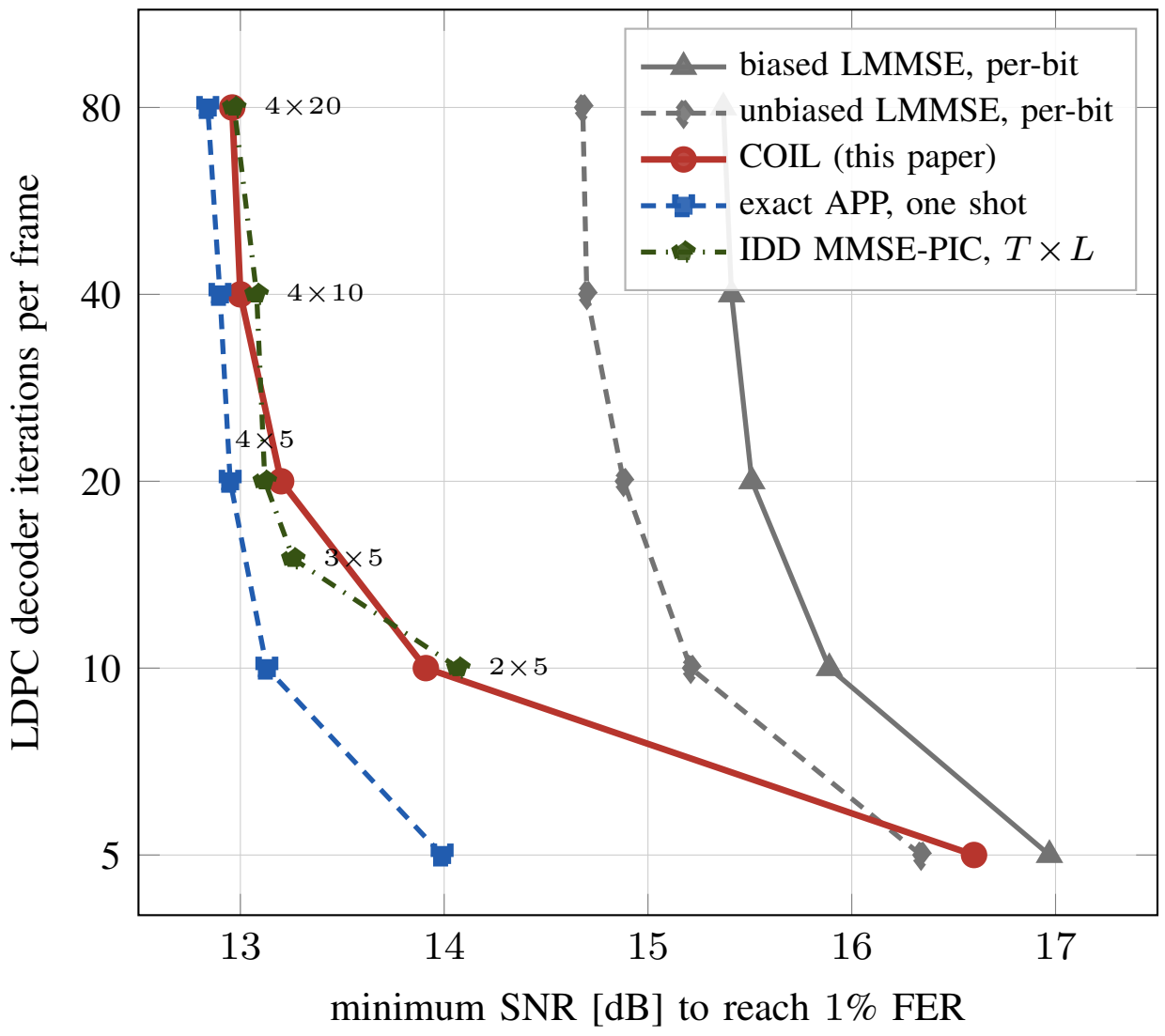
\begin{figure}[tp]
\centering
\vspace{0.22cm} % manual fix
\begin{tikzpicture}
\begin{semilogyaxis}[
  scale only axis, width=0.81\columnwidth, height=0.72\columnwidth,
  log basis y={2},
  xlabel={minimum SNR [dB] to reach $1$\% FER},
  ylabel={LDPC decoder iterations per frame},
  xmin=12.5, xmax=17.5, ymin=4, ymax=115,  
  ytick={5,10,20,40,80},
  yticklabels={5,10,20,40,80},
  grid=both,
  grid style={line width=0.2pt, draw=black!12},
  major grid style={line width=0.2pt, draw=black!20},
  tick label style={font=\footnotesize},
  label style={font=\footnotesize},
  legend style={font=\scriptsize, at={(0.98,0.98)}, anchor=north east,
                draw=black!30, fill=white, fill opacity=0.9,
                text opacity=1, row sep=-1pt, inner sep=2.5pt},
  legend cell align={left},
]
\addplot[biaStyle]  coordinates {(16.97,5) (15.89,10) (15.51,20) (15.41,40) (15.37,80)};
  \addlegendentry{biased LMMSE, per-bit} 
\addplot[unbStyle]  coordinates {(16.34,5) (15.21,10) (14.88,20) (14.70,40) (14.68,80)};
  \addlegendentry{unbiased LMMSE, per-bit}  
\addplot[coilStyle] coordinates {(16.60,5) (13.91,10) (13.20,20) (13.00,40) (12.96,80)};  
  \addlegendentry{COIL (this paper)}
\addplot[appStyle]  coordinates {(13.99,5) (13.13,10) (12.95,20) (12.90,40) (12.84,80)};
  \addlegendentry{exact APP, one shot}
\addplot[iddStyle]  coordinates {(14.07,10) (13.26,15) (13.12,20) (13.08,40) (12.97,80)};  
  \addlegendentry{IDD MMSE-PIC, $T\!\times\!L$}
\node[font=\tiny,anchor=west]  at (axis cs:14.13,10)  {$2{\times}5$};
\node[font=\tiny,anchor=west]  at (axis cs:13.32,15)  {$3{\times}5$};
\node[font=\tiny,anchor=south,yshift=2.5pt] at (axis cs:13.12,20) {$4{\times}5$};
\node[font=\tiny,anchor=west]  at (axis cs:13.14,40)  {$4{\times}10$};
\node[font=\tiny,anchor=west]  at (axis cs:13.03,80)  {$4{\times}20$};
\end{semilogyaxis}
\end{tikzpicture}
\caption{Performance--complexity trade-off showing the minimum required SNR to reach a target FER of $1\%$ and the associated total number of LDPC decoding iterations (bottom left is better).
Colors and markers identify the receiver (cf.~\fref{fig:fer}), and each IDD point is labeled $T{\times}L$, i.e., $T$ outer loops with $L$ decoder iterations (inner loops) each.
COIL closely approaches the exact one-shot APP detector and IDD with MMSE-PIC as the number of LDPC decoder iterations increases.}
\label{fig:tradeoff}  
\end{figure}

%%%
\subsection{Performance--Complexity Trade-Off}
\label{sec:results-tradeoff}

We now evaluate the performance--complexity trade-off between the minimum SNR required to achieve an FER of $1\%$ and the number of layered LDPC decoder iterations.
We only measure the number of LDPC decoder iterations and exclude the complexity of the data detector and COIL's additional coupling-check updates (cf.~\fref{sec:props}).
Excluding detector complexity favors IDD because COIL runs the data detector once, whereas IDD re-runs it $T$ times.
We see in \fref{fig:tradeoff}\footnote{All SNR operating points, including those in \fref{fig:tradeoff}, are obtained by linear interpolation from FER simulations in the log domain; thus those SNR values deviate slightly from visually inspecting the FER curves (e.g.,~\fref{fig:fer}).} that COIL closely approaches the one-shot APP detector as the decoder complexity grows, to within $0.10$\,dB at $40$ decoder iterations.
We also see that one-shot COIL at $40$ decoder iterations matches the performance of four-iteration IDD with the MMSE-PIC algorithm.
This result is particularly interesting because we compare the same \emph{total} number of LDPC decoder iterations: IDD distributes its $40$ iterations across $T=4$ outer iterations with $L=10$ decoder iterations each, whereas COIL uses all $40$ iterations in a single decoder run. Thus, COIL achieves comparable performance even without accounting for the complexity of re-running the data detector within IDD.
With only five LDPC decoder iterations, however, COIL requires more SNR than the conventional one-shot receiver with an unbiased LMMSE detector.

%%%
\subsection{SNR Operating Point for Different System Configurations}
\label{sec:results-config}
\fref{tab:configs} compares the SNR operating point, i.e., the minimum SNR required to achieve a target FER of $1\%$, across antenna configurations (written as $\MR{\times}\MT$), modulation schemes, code rates, and code lengths. We simulate $20$ flooding iterations of the LDPC decoder, which also demonstrates that COIL works with a flooding schedule.
We report the SNR operating points for the conventional per-bit interface, COIL based on the biased LMMSE summary, the one-shot exact APP benchmark, and four-iteration IDD with MMSE-PIC. We also report the difference~$\Delta$ between the per-bit and COIL operating points.
The per-bit column is the unbiased LMMSE receiver, which is the most widely used equalization-based one-shot receiver that keeps the conventional interface and which coincides with the $T=1$ point of MMSE-PIC.
For the $8\times4$ and the long-code $4\times4$ configurations, one-shot COIL operates within $0.1$\,dB of the exact one-shot APP detector. For the $4\times4$ rate-$3/4$ configuration with 16-QAM, the gap is larger at $1.2$\,dB. Exact APP results are unavailable for the 64-QAM and $8\times8$ configurations; dashes in \fref{tab:configs} mark the unavailable results.
The comparison with IDD depends on the system configuration. For example, COIL performs better than four-iteration IDD in both rate-$3/4$ configurations, where the SNR gain of COIL over the per-bit receiver ($4.0$\,dB and $4.1$\,dB for 64-QAM and 16-QAM, respectively) exceeds that of IDD ($2.0$\,dB and $2.9$\,dB).
As presented, COIL is a one-shot receiver whose advantage is largest when a data-detector pass is expensive relative to a decoder iteration.

\begin{table}[tp]
\centering
\caption{Minimum SNR [dB] required to achieve a target FER of $1\%$ \\ for different system configurations.}
\label{tab:configs}
\footnotesize
\setlength{\tabcolsep}{4.5pt}
\begin{tabular}{@{}lccccc@{}}
\toprule
& \multicolumn{4}{c}{One shot} & IDD \\
\cmidrule(lr){2-5} \cmidrule(l){6-6}
Configuration & Per-bit & COIL & $\Delta$ & Exact APP & $T{=}4$ \\
\midrule
$8{\times}4$, 16-QAM  &  7.1 &  6.9 & 0.2 &  6.8 &  6.1 \\
$4{\times}4$, 16-QAM, $\Ncw{=}1944$ & {14.8} & {12.9} & {1.9} & {12.8} & {12.4} \\
$8{\times}8$, 16-QAM  & {15.5} & {13.9} & {1.6} & ---   & {12.8} \\
$4{\times}4$, 64-QAM, $R{=}3/4$  & {28.9} & {24.9} & {4.0} & ---   & {26.9} \\
$4{\times}4$, 16-QAM, $R{=}3/4$  & {22.4} & {18.3} & {4.1} & {17.1} & {19.5} \\
\bottomrule
\end{tabular}
\end{table}

%%%
\subsection{Cost of Approximate Coupling-Logit Computation}
\label{sec:results-cost}
We finally compare the performance of the various tree constructions discussed in \fref{sec:treeconstr} and \fref{sec:cheapweight}.
To this end, we report the SNR operating point, i.e., the minimum SNR required to achieve a tree-metric rate $R_{\up{tree}}$ of $8$ bits, for a $\MT=\MR=4$ system with 16-QAM and a biased LMMSE front end.
We also report the loss with respect to the benchmark: tree scores and coupling logits computed from the exact Gaussian bit-pair metrics.
\fref{tab:cheapcost} supports the recommendation of \fref{rem:selectcheap}: use~\fref{eq:graycheap} for tree construction and exact coupling logits~\fref{eq:couplinglogit}. This approach loses only $0.08$\,dB relative to the exact method while avoiding a quadratic number of expensive marginalizations.
Using the max-log approximation for both tree construction and coupling-logit computation instead incurs a $0.35$\,dB SNR loss.

\begin{table}[tp]
\centering
\caption{Minimum SNR [dB] required to achieve a tree-metric rate \\ of $R_{\up{tree}}=8$\,bits for various COIL computation methods.}
\label{tab:cheapcost}
\footnotesize
\setlength{\tabcolsep}{4pt}
\begin{tabular}{@{}llcc@{}}
\toprule
Coupling logits & Tree construction & SNR [dB] & Loss [dB] \\
\midrule
Exact~\fref{eq:pairmarginal} & Exact pairs~\fref{eq:edgeweight}  & 9.08  & ---  \\
Exact~\fref{eq:pairmarginal} & Gray~\fref{eq:graycheap}  & 9.16  & 0.08 \\
Max-log~\fref{eq:maxlog}  & Max-log~\fref{eq:maxlog}   & 9.43  & 0.35 \\
Max-log~\fref{eq:maxlog}  & Gray~\fref{eq:graycheap}   & 9.58  & 0.50 \\
Gray~\fref{eq:graycheap}   & Gray~\fref{eq:graycheap}    & 10.03 & 0.95 \\
\bottomrule
\end{tabular}
\end{table}

%%%%%%%%%%%%%%%%%%%%%%%%%%%%%%%%%%%%%%%%%%%%%%%%%%%%%%%%%%%
%% Limitations

\section{Limitations and Open Problems}
\label{sec:limitations}
We now discuss the limitations of our results and possible directions for further work.

%-----------------------------------------------------------
\subsection{Limitations}
\label{sec:limits}

Our results have the following limitations.

\subsubsection{Empirical Scaling}
In all simulations, we empirically scale the soft inputs as described in \fref{rem:scaling} for each receiver and system configuration. We currently lack theory for selecting this scaling parameter, and a poor choice can cause a performance loss that is larger than the gain from COIL.

\subsubsection{Reliance on Simulation}
All of our rate comparisons and FER results rely on Monte-Carlo simulations. 
Deriving closed-form expressions appears difficult: even the conditional total correlation in~\fref{eq:bicmloss} depends on the channel statistics, constellation, and labeling. Similarly, the finite-size factor graphs complicate an analytical characterization of the FER.

\subsubsection{Interface Bounds Are Not Receiver Bounds}
With uniform i.i.d.\ bits, exact one-shot APP logits maximize the GMI of the per-bit product metric. This rate benchmark is not an FER bound for a finite-length detector--decoder receiver.
For example, a soft-input soft-output variant of the same data detector in an IDD receiver might reach lower SNR operating points, although such a receiver would be impractical.

\subsubsection{Restricted Class of Detectors}
\fref{defi:eqdet} covers equalization-based data detectors, and we only consider ZF, (biased and unbiased) LMMSE, and the exact APP detector. 
Whether other detectors that do not report Gaussian summaries can produce useful (and even better) coupling logits remains untested.

\subsubsection{Dynamic Range of Coupling Logits}
The magnitudes of the coupling logits can grow without bound as the SNR increases. Thus, any fixed-point implementation of a COIL receiver must apply magnitude clipping.
Our floating-point simulations reveal that far above the shown operating points, the FER of COIL increases with SNR.
Thus, the dynamic range of the coupling logits will require careful treatment in practical receivers that rely on fixed-point arithmetic.

%-----------------------------------------------------------
\subsection{Open Problems}
\label{sec:open}

Further work could address the following questions.

\subsubsection{Optimal COIL Receiver}
We do not know the best equalizer for computing COIL soft outputs. 
Our own simulations have shown that using ZF instead of LMMSE equalization with COIL performs poorly, even though the ZF summary is information-lossless at the vector level.
Furthermore, the unbiased LMMSE equalizer improves the per-bit interface but degrades $R_{\up{tree}}$.
The equalizer must provide useful per-bit soft outputs and coupling logits; however, the choice that maximizes~$R_{\up{tree}}$ is unknown.

\subsubsection{Low-Complexity Coupling Logits}
In \fref{sec:cheapweight}, we have shown how to reduce the complexity of tree construction. However, computing the $N-1$ coupling logits still requires marginalization over $|\setO|^2$ symbol pairs. An open question is whether an approximation can reduce this cost while retaining the accuracy of exact marginalization.

\subsubsection{Other Bit Pairs and Labelings}
We have restricted our analysis and coupling-logit computation to candidate pairs in different streams. We have not investigated intra-symbol coupling logits, possibly with different bit labelings.

\subsubsection{Hybrid COIL-IDD Receivers}
Nothing in \fref{sec:coil} requires us to compute the COIL soft outputs only once. 
One could, for example, design an MMSE-PIC-based receiver that computes COIL soft outputs in every outer iteration. 
This approach would yield a hybrid COIL-IDD receiver architecture whose performance and trade-offs remain unexplored.

\subsubsection{Hardware Prototype}
COIL increases the complexity of the data detector and of the channel decoder, and it increases memory requirements; \fref{sec:props} quantifies these increases in terms of metric evaluations, boxplus operations, and number of computed (and stored) soft values.
Nonetheless, a hardware implementation of a full COIL receiver is required to determine its actual complexity, memory, latency, and power trade-offs with respect to IDD implementations.

%-----------------------------------------------------------
\section{Conclusions}
\label{sec:conclusions}

We have shown that one-shot MIMO-BICM receivers are limited primarily by their per-bit soft-output interface rather than by the used data detector.
For a given decoding metric, the rate loss is the minimum expected KL divergence from the true posterior to its normalized, scaled versions.
Our experiments have shown that COIL recovers a large fraction of this loss using one coupling logit per edge of a spanning tree. With exact posterior marginals, the tree metric recovers exactly the conditional mutual information of the bit pairs connected by its edge set.
We have also shown that a conventional belief-propagation decoder can process this format using one extra degree-three check node per coupling logit.
Simulation results have demonstrated that one-shot COIL receivers can approach the error-rate performance of IDD receivers, while avoiding (i) the repeated MIMO data-detector passes that result in high complexity and latency, and (ii) complicated scheduling between the detector and decoder stages~\cite{studer09}.

We believe that some of the presented results are of interest beyond the studied MIMO-BICM application. 
Our results show that a detector--decoder interface need not be limited to one soft value per coded bit.
Specifically, COIL demonstrates that a data detector can report a small amount of additional information, e.g., one number per selected bit pair, and that a conventional channel decoder can process this information without changing its decoding algorithm.
This observation motivates the study of richer (or alternative) soft-output formats in other receivers that separate detection from decoding.

%%%%%%%%%%%%%%%%%%%%%%%%%%%%%%%%%%%%%%%%%%%%%%%%%%%%%%%%%%%
%% Appendix: Proofs
%%%%%%%%%%%%%%%%%%%%%%%%%%%%%%%%%%%%%%%%%%%%%%%%%%%%%%%%%%%

\appendix[Proofs] 

%%%
\subsection{Proof of Achievability in Definition~\ref{defi:gmi}} 
\label{app:gmi}
The mismatched-decoding GMI~\cite{ganti00,martinez09} permits arbitrary discrete or continuous alphabets. Here, the finite input alphabet~$\setA$ contains one transmit vector or bit pattern per resource block. The output is the observed continuous pair~$\{\vec{y},\mat{H}\}$ used to evaluate the metric.
With i.i.d.\ codebooks drawn uniformly from $\setA$, the standard random-coding argument establishes that every rate below~\fref{eq:gmi} is achievable by a decoder that maximizes $\prod_{\kappa=1}^{K}q^{(\kappa)}(\vec{x}^{(\kappa)})$ over all codewords; the superscript $(\kappa)$ indexes the $K$ resource blocks.
This decoder does not depend on $\tau$, which enters only as a parameter in the achievability bound; every $\tau\geq0$ yields an achievable rate, and the largest rate is obtained at the maximizer~$\tau^\star$.
Note that independence of the resource blocks is not required: the random-coding argument needs only the per-codeword empirical metric averages to converge to the expectations in~\fref{eq:gmi}, which holds for stationary, ergodic channel matrices with independent i.i.d.\ inputs and noise, provided that the per-resource-block log-ratios in~\fref{eq:gmi} have finite absolute mean.%
\footnote{For i.i.d.\ resource blocks, $R(q)$ in~\fref{eq:gmi} is the standard GMI of~\cite{kaplan93,ganti00,martinez09}; the necessary regularity conditions are provided in~\cite{scarlett20}. For ergodic resource blocks, we invoke the channels-with-memory result of~\cite[Sec.~IV]{ganti00}, which is stated for general input and output sequences. In  our setting, the ergodic theorem replaces the law of large numbers.}
Furthermore, no assumption on the accuracy of the metric $q$ enters this random-coding argument, i.e., mismatch only lowers the rate~$R(q)$
of~\fref{eq:gmi}.

%%%
\subsection{Proof of Lemma~\ref{lem:divergence}} 
\label{app:divergence}
We first prove statement~(i). The quantity inside the expectation
in~\fref{eq:gmi} is $\log_2(\bar{q}_\tau(\vec{x}))$. We evaluate its expectation
in two steps, first over $\vec{x}$ for a fixed $\vec{y}$, then over $\vec{y}$.
For a fixed $\vec{y}$, the transmitted input is distributed as
$p(\cdot\,|\,\vec{y})$, and adding and subtracting $\log_2(p(\vec{x}\,|\,\vec{y}))$
splits the average into an entropy and a divergence, i.e., we have
\begin{align}\label{eq:kldecomp}
\Ex_{p(\cdot|\vec{y})}\!\big[\log_2(\bar{q}_\tau)\big] = -H\big(p(\cdot\,|\,\vec{y})\big) - \mathrm{KL}\big(p(\cdot\,|\,\vec{y})\,\big\|\,\bar{q}_\tau\big).  
\end{align}
Averaging~\fref{eq:kldecomp} over $\vec{y}$ and adding $N$ yields~\fref{eq:divergence}, as we show next.
First, the entropy term in~\fref{eq:kldecomp} averages to $-H(\vec{x}\,|\,\vec{y})$.
Second, the coded bits are uniform i.i.d.\ (cf.\ \fref{sec:model}), so the added constant is $N=H(\vec{x})$.
Third, adding $N=H(\vec{x})$ to $-H(\vec{x}\,|\,\vec{y})$ yields $H(\vec{x})-H(\vec{x}\,|\,\vec{y})=I(\vec{x};\vec{y})$.
Since the bit labeling is a bijection, $\vec{x}$ and~$\vec{s}$ determine each other, and $I(\vec{x};\vec{y})=I(\vec{s};\vec{y})$, which is the first term of~\fref{eq:divergence}.
Since the KL divergence is nonnegative, we have $R(q)\leq I(\vec{s};\vec{y})$, with equality if and only if the two arguments of the KL divergence are equal; this is the equality condition.

We now prove statement~(ii), for which we only need to show that $\tau=1$ maximizes $R_\tau(q)$; the rate then follows by inserting $\tau=1$ into~\fref{eq:divergence}. 
The function $R_\tau(q)$ is concave in $\tau$: the first term of $\log_2(\bar q_\tau)$ is linear in~$\tau$, whereas the second is the negative of a log-sum-exp function of~$\tau$ and is therefore concave.
We now show that the derivative is zero at $\tau=1$:
\begin{align} \label{eq:gmideriv}
\frac{\partial R_\tau(q)}{\partial\tau} = \frac{1}{\ln(2)}\,\Ex\!\left[\Ex_{p(\cdot|\vec{y})}[\ln(q)]-\Ex_{\bar{q}_\tau}[\ln(q)]\right]
\end{align}
is the average of $\ln(q)$ evaluated at the transmitted input~$\vec{x}$, minus its average when the input is instead drawn from~$\bar{q}_\tau$.
The outer expectation is over the receive vector and the channel realizations, as in~\fref{eq:gmi}.
Since $q$ is a PMF, we have $\bar{q}_1=q$, so the assumption in statement~(ii) causes the expectation in~\fref{eq:gmideriv} to be zero at~\mbox{$\tau=1$}.

%%%
\subsection{Proof of Proposition~\ref{prop:natural}} 
\label{app:natural}
For every $\vec{s}\in\setO^{\MT}$, we have
\begin{align}
\|\vec{y}-\mat{H}\vec{s}\|_2^2   = \|\vec{y}-\mat{H}\hat{\vec{s}}_{\up{ZF}}\|_2^2   + (\vec{s}-\hat{\vec{s}}_{\up{ZF}})\herm\mat{G}\,(\vec{s}-\hat{\vec{s}}_{\up{ZF}}).
\end{align}
The first term does not depend on $\vec{s}$ and is absorbed by the normalization in~\fref{eq:posterior}. Dividing by $\Nzero$ yields
\begin{align}  \label{eq:natexp}
p(\vec{s}\,|\,\vec{y}) \propto\e^{-(\vec{s}-\hat{\vec{s}}_{\up{ZF}})\herm\mat{G}\,(\vec{s}-\hat{\vec{s}}_{\up{ZF}})/\Nzero}.
\end{align}
It remains to show that~\fref{eq:natexp} is equal to~\fref{eq:vecmetric}.
The ZF error is proper (cf.\ \fref{rem:zfexact}), and for a proper covariance matrix~$\hat{\mat{C}}$ the real-valued decomposition~\fref{eq:realdecomp} preserves quadratic forms, $\tfrac12\vec{v}_\Reals\trans\hat{\mat{C}}_\Reals^{-1}\vec{v}_\Reals=\vec{v}\herm\hat{\mat{C}}^{-1}\vec{v}$ for every $\vec{v} \in\Complex^{\MT}$.
Applying this result to $\hat{\mat{C}}_{\up{ZF}}=\Nzero\mat{G}^{-1}$, with inverse $\mat{G}/\Nzero$, turns~\fref{eq:vecmetric} into~\fref{eq:natexp}, which proves~\fref{eq:natkernel}.

For the rate, we note that~\fref{eq:natkernel} is the equality condition of \fref{lem:divergence}(i) for $\tau=1$.
Setting $\tau=1$ in \fref{eq:pmfthingie} makes $\bar{q}_1$ the normalized metric, which by~\fref{eq:natkernel} \emph{is} the posterior.
Hence, the KL divergence in~\fref{eq:divergence} vanishes and we obtain $R_{\up{vec}}=R_1(q_{\up{vec}})=I(\vec{s};\vec{y})$.

%%%
\subsection{Proof of Corollary~\ref{cor:tilt}} 
\label{app:tilt}
Since the assumed error model is proper, the quadratic-form identity in the proof of \fref{prop:natural} applies, and the exponent of the LMMSE metric~\fref{eq:vecmetric} with $\hat{\mat{C}}_{\up{L}}=\Nzero\mat{A}^{-1}$ from~\fref{eq:lmmse} corresponds to the complex-valued quadratic form $(\vec{s}-\hat{\vec{s}}_{\up{L}})\herm \mat{A} (\vec{s}-\hat{\vec{s}}_{\up{L}})/\Nzero$.
We have the following equality:
\begin{align}
(\vec{s}-\hat{\vec{s}}_{\up{L}})\herm \mat{A} (\vec{s}-\hat{\vec{s}}_{\up{L}})= \vec{s}\herm\mat{A}\vec{s} \!-\!2 \Re(\vec{s}\herm\mat{A}\hat{\vec{s}}_{\up{L}})+ \kappa(\vec{y}).
\end{align}
Here, $\kappa(\vec{y})\eqdef \hat{\vec{s}}_{\up{L}}\herm\mat{A} \hat{\vec{s}}_{\up{L}}$ collects all terms that do not depend on $\vec{s}$.
With $\mat{A}\hat{\vec{s}}_{\up{L}}=\mat{H}\herm\vec{y}$ and $\vec{s}\herm\mat{A}\vec{s}=\vec{s}\herm\mat{G}\vec{s} +\tfrac{\Nzero}{\Es}\|\vec{s}\|_2^2$, the exponent corresponds to $\tfrac{1}{\Nzero}\big(\|\vec{y}-\mat{H}\vec{s}\|_2^2-\|\vec{y}\|_2^2\big) +\tfrac{1}{\Es}\|\vec{s}\|_2^2+\tfrac{1}{\Nzero}\kappa(\vec{y})$, which proves~\fref{eq:tilt}. 
For constant-modulus constellations, we have $\|\vec{s}\|_2^2=\MT\Es$ for every $\vec{s}\in\setO^{\MT}$. Thus, the metric is proportional to the posterior, and \fref{lem:divergence}(i) establishes equality for $\tau=1$.

For the strict inequality $R_{\up{vec}}<I(\vec{s};\vec{y})$, \fref{lem:divergence}(i) requires us to rule out \emph{every} $\tau\geq0$.
Raising~\fref{eq:tilt} to the power $\tau$ yields $q_{\up{vec}}^\tau\propto p(\vec{s}\,|\,\vec{y})^{\tau}\e^{-\tau\|\vec{s}\|_2^2/\Es}$, so this metric is proportional to the posterior if and only if~$p(\vec{s}\,|\,\vec{y})^{\tau-1}\e^{-\tau\|\vec{s}\|_2^2/\Es}$ does not depend on~$\vec{s}$. 
Taking the negative logarithm of this expression and substituting the explicit posterior~\fref{eq:posterior} for $p(\vec{s}\,|\,\vec{y})$ shows that
\begin{align}  \label{eq:tiltcond}
\tfrac{\tau-1}{\Nzero}\|\vec{y}-\mat{H}\vec{s}\|_2^2+ \tfrac{\tau}{\Es}\|\vec{s}\|_2^2
\end{align}
must be constant for all $\vec{s}\in\setO^{\MT}$. 
In~\fref{eq:tiltcond}, $\vec{s}$ and $\vec{y}$ only appear jointly in $-\tfrac{2(\tau-1)}{\Nzero}\Re(\vec{s}\herm\mat{H}\herm\vec{y})$, and this term varies with $\vec{s}$ unless $\tau=1$. Setting $\tau=1$ results in $\|\vec{s}\|_2^2/\Es$, which is the same for all $\vec{s}\in\setO^{\MT}$ only for constant-modulus constellations; the case $\tau=0$ follows the same argument. 
Hence, no $\tau\geq0$ achieves equality.
So far, we have shown that the divergence $\Ex[\mathrm{KL}(p(\cdot\,|\,\vec{y})\|\bar{q}_\tau)]$ in~\fref{eq:divergence} is positive for every fixed $\tau\geq0$.
As \fref{lem:divergence}(i) subtracts the minimum of this divergence over~$\tau$, it could still be zero if the divergence approached zero as~$\tau$ increases.
However, the divergence is continuous in~$\tau$ and grows without bound as $\tau\to\infty$. Its minimum over $\tau\geq0$ is therefore attained at a finite~$\tau$, where the divergence is positive.
We conclude that $R_{\up{vec}}<I(\vec{s};\vec{y})$.

%%%
\subsection{Proof of Proposition~\ref{prop:bicmloss}} 
\label{app:bicmloss}
With exact logits, each per-bit metric is the true posterior marginal $q_k(x_k)=p(x_k\,|\,\vec{y})$, so the product metric is $q_{\up{bit}}=\prod_{k=1}^{N}p(x_k\,|\,\vec{y})$. 
This is a PMF on $\{\pm1\}^N$ and has the same per-bit marginals as $p(\vec{x}\,|\,\vec{y})$. Since $\ln(q_{\up{bit}})$ is a sum of $N$ functions of single bits, its mean depends only on those marginals and is therefore the same under $p(\cdot\,|\,\vec{y})$ and under $q_{\up{bit}}$. 
This is exactly the hypothesis of \fref{lem:divergence}(ii), which yields $\tau^\star=1$ and $R_{\up{bit}}=I(\vec{s};\vec{y})-\Ex[\mathrm{KL}(p(\cdot\,|\,\vec{y})\|q_{\up{bit}})]$. 
Expanding the divergence using this product of marginals gives
\begin{align}
  \Ex\!\left[\mathrm{KL}\big(p(\cdot\,|\,\vec{y})\,\big\|\,q_{\up{bit}}\big)\right]
  & = \sum_{k=1}^{N}H(x_k\,|\,\vec{y})-H(\vec{x}\,|\,\vec{y}),
  \label{eq:bitkl}
\end{align}
which corresponds to~\fref{eq:bicmloss}.
Alternatively, instead of expanding the divergence, one can evaluate~\fref{eq:bitgmi} directly at $\tau=1$; this gives the equivalent form $R_{\up{bit}}=\sum_{k=1}^{N}I(x_k;\vec{y})$ used in \fref{rem:sparam}, since its $k$th summand is $1-H(x_k\,|\,\vec{y})=I(x_k;\vec{y})$.

%%%
\subsection{Proof of Lemma~\ref{lem:ising}} 
\label{app:ising}
Any strictly positive function of $(x_k,x_l)\in\{\pm1\}^2$ can be written as $\e^{\alpha+h_k x_k+h_l x_l+\gamma x_k x_l/2}$ with four real
coefficients, because the four monomials $1$, $x_k$, $x_l$, and $x_kx_l$ span all real-valued functions on the four sign combinations.
Solving this system for the four unknowns yields $\gamma=\gamma_{k,l}$ as in~\fref{eq:couplinglogit}, $2h_k=\tilde{\lambda}_k+\delta_k^{(k,l)}$ as in~\fref{eq:field}, and $2h_l=\tilde{\lambda}_l+\delta_l^{(k,l)}$ as in~\fref{eq:fieldl}, i.e., the Ising fields~\fref{eq:isingfields}. 
Dividing by $q_k(x_k)\>q_l(x_l)\propto\e^{\tilde{\lambda}_k x_k/2+\tilde{\lambda}_l x_l/2}$ subtracts $\tilde{\lambda}_k/2$ and $\tilde{\lambda}_l/2$ from the linear coefficients, which yields~\fref{eq:isingfactor}; we absorb the constant~$\alpha$ in the proportionality factor. 
Since none of these steps used any property of $q_k$ beyond positivity, the identity holds for arbitrary positive per-bit metrics. Finally, \fref{eq:couplinglogit}--\fref{eq:fieldl} are all log-ratios with the same number of $q_{k,l}$ terms in the numerator and denominator; thus, constant scaling of $q_{k,l}$ is irrelevant.

%%%
\subsection{Proof of Lemma~\ref{lem:cheapgamma}} 
\label{app:cheapgamma}
Write the exponent of~\fref{eq:pairkernel} as $g(\vec{u})=-\frac12\vec{u}\trans\mat{P}\vec{u}$ with $\vec{u}=\vec{v}_\Reals-\vec{m}_\Reals$. Here, $\vec{v}_\Reals$ represents the nearest symbol pair or one of its flipped versions in \fref{lem:cheapgamma}, and $\vec{m}_\Reals$ is the two-stream mean.
Use the reordered coordinates of~\fref{eq:precblocks} and write $\vec{u}=[\vec{u}_i;\vec{u}_j]$, so that
\begin{align} 
g(\vec{u}) = \textstyle -\frac12\vec{u}_i\trans\mat{P}_{ii}\vec{u}_i - \frac12\vec{u}_j\trans\mat{P}_{jj}\vec{u}_j - \vec{u}_i\trans\mat{P}_{ij}\vec{u}_j.
\end{align}
By~\fref{eq:couplinglogit}, the coupling logit is half the signed sum of the four values of $\ln(q_{k,l})$: hypotheses for which $x_k$ and $x_l$ agree have a plus sign; the others have a minus sign.
Under the approximation~\fref{eq:maxlog}, these four values are evaluations of $g$ at the nearest symbol pair and its flipped versions: flipping bit~$k$ shifts~$\vec{u}_i$ by~$\vec{\alpha}_k$, and flipping bit~$l$ shifts~$\vec{u}_j$ by~$\vec{\beta}_l$.
In this signed sum, the two quadratic terms of $g$ cancel, and the term $-\frac12\vec{u}_i\trans\mat{P}_{ii}\vec{u}_i$ depends on $x_k$ only. Thus, each of its two values occurs once with a plus and once with a minus, and the same argument applies to the term in $\vec{u}_j$.
Only the bilinear term remains, and it contributes $-\vec{\alpha}_k\trans\mat{P}_{ij}\vec{\beta}_l$ regardless of the value of~$\vec{u}$.
It remains to determine the appropriate signs.
The signed sum above adds the two hypotheses in which the nearest symbol pair is flipped in both streams or in neither, and subtracts the two in which it is flipped in exactly one stream.
This is the combination of~\fref{eq:couplinglogit} if $\hat{x}_k\hat{x}_l=+1$ and its negated version if $\hat{x}_k\hat{x}_l=-1$, which is precisely the factor $\hat{x}_k\hat{x}_l$ in~\fref{eq:cheapgamma}; the remaining $\frac12$ is the prefactor in~\fref{eq:couplinglogit}.

%%%
\subsection{Proof of Proposition~\ref{prop:oracletree}} 
\label{app:oracletree}
Since $\Nzero>0$, the posterior~\fref{eq:posterior} has full support, so all of its marginals are strictly positive and the tree metric is a valid decoding metric in the sense of \fref{defi:gmi}.
With exact marginals, the tree metric in~\fref{eq:treemetric} corresponds to the classical tree approximation by Chow and Liu~\cite[Eq.~1]{chowliu68} applied to $p(\vec{x}\,|\,\vec{y})$, and we will use the following two properties.

\begin{prty} \label{prty:pmf}
The tree metric $q_{\up{tree}}$ in~\fref{eq:treemetric}, built from the exact posterior marginals, is a PMF on $\{\pm1\}^N$.
\end{prty}
To show \fref{prty:pmf}, we marginalize~\fref{eq:treemetric} over one bit at a time.
Pick a bit $x_l$ contained in exactly one edge $(k,l)\in\setE$; every tree (or forest) with at least one edge contains such a bit.
The only factors of~\fref{eq:treemetric} that contain $x_l$ are the per-bit metric $q_l(x_l)$ and the correction factor $c_{k,l}(x_k,x_l)$. Their product is $q_{k,l}(x_k,x_l)/q_k(x_k)$.
Since the pair marginal satisfies $\sum_{x_l}q_{k,l}(x_k,x_l)=q_k(x_k)$, marginalizing this product over $x_l\in\{\pm1\}$ yields one.
What remains is the tree metric associated with the bits and edges that are left after removing the bit $x_l$ and the edge $(k,l)$.
Repeating this argument until no edge is left, and then marginalizing the remaining per-bit metrics, which are PMFs themselves, yields $\sum_{\vec{x}\in\{\pm1\}^N}q_{\up{tree}}(\vec{x})=1$.

\begin{prty} \label{prty:marginals}
The tree metric $q_{\up{tree}}$ has the same marginals as the posterior $p(\cdot\,|\,\vec{y})$ for every single bit and for every bit pair that is connected by an edge of~$\setE$. 
\end{prty}
To show \fref{prty:marginals}, we reuse the marginalization argument of \fref{prty:pmf}, but exclude the bits of interest from the marginalization.
Marginalizing $q_{\up{tree}}$ over all bits except $x_m$ yields one at every step. The last remaining factor is $q_m(x_m)=\Pr[x_m\,|\,\vec{y}]$, i.e., the posterior marginal of that bit.
Marginalizing over all bits except the two bits of one edge $(k,l)\in\setE$ leaves the factors $q_k(x_k)\,q_l(x_l)\,c_{k,l}(x_k,x_l)=q_{k,l}(x_k,x_l)=\Pr[x_k,x_l\,|\,\vec{y}]$, i.e., the posterior marginal of that bit pair.
We emphasize that \fref{prty:marginals} makes no statement about bit pairs that $\setE$ does not~connect.

With \fref{prty:pmf} and \fref{prty:marginals}, we can prove the proposition in the following four steps.

\subsubsection*{Step 1} 
The logarithm $\ln(q_{\up{tree}})$ is a sum of terms, each of which depends either on a single bit or on the two bits of one edge of~$\setE$.
Therefore, its mean depends on the underlying distribution only through the marginals of these bits and bit pairs, and \fref{prty:marginals} states that $q_{\up{tree}}$ and the posterior share exactly these marginals.
The two means thus coincide, which is the hypothesis of \fref{lem:divergence}(ii), and we conclude that $\tau^\star=1$.

\subsubsection*{Step 2}
Taking the logarithm of~\fref{eq:treemetric} with exact marginals yields
\begin{align} \label{eq:treelog}
\log_2\big(q_{\up{tree}}(\vec{x})\big) =\, &  \sum_{k=1}^{N}\log_2\big(\Pr[x_k\,|\,\vec{y}]\big) \notag \\
& + \!\!\!\sum_{(k,l)\in\setE}\!\!\!\log_2\!\left(\frac{\Pr[x_k,x_l\,|\,\vec{y}]}{\Pr[x_k\,|\,\vec{y}]\Pr[x_l\,|\,\vec{y}]}\right)\!.
\end{align}
We now take the expectation of~\fref{eq:treelog} with respect to $\vec{x}\sim p(\cdot\,|\,\vec{y})$ for a fixed receive vector~$\vec{y}$.
The first sum yields negative per-bit entropies, and the second yields pairwise MIs, because the logarithms contain the posterior's own marginals.
Using $\mathrm{KL}(p(\cdot\,|\,\vec{y})\|q_{\up{tree}}) = -H\big(p(\cdot\,|\,\vec{y})\big)-\Ex_{p(\cdot|\vec{y})}[\log_2(q_{\up{tree}})]$ and averaging over the receive vectors~$\vec{y}$ results in
\begin{align}
  &\Ex\!\left[\mathrm{KL}\big(p(\cdot\,|\,\vec{y})\,\big\|\,q_{\up{tree}}\big)\right] \nonumber\\
  &\quad = \sum_{k=1}^{N}H(x_k\,|\,\vec{y})-H(\vec{x}\,|\,\vec{y})
  -\!\!\!\sum_{(k,l)\in\setE}\!\!\! I(x_k;x_l\,|\,\vec{y}),
  \label{eq:treekl}
\end{align}
i.e., the conditional total correlation of~\fref{eq:bicmloss} reduced by the pairwise MI of the bit pairs that $\setE$ connects.

\subsubsection*{Step 3}
Inserting~\fref{eq:treekl} into \fref{lem:divergence}(ii) and subtracting~\fref{eq:bicmloss} results in~\fref{eq:oracletree}.

\subsubsection*{Step 4}
Every step so far was carried out for a fixed $\vec{y}$; in particular, the hypothesis of \fref{lem:divergence}(ii) is a condition per receive vector.
Therefore, choosing $\setE$ as a function of $\vec{y}$ changes nothing except that the edge sum stays inside the expectation, which yields~\fref{eq:oracletreey}.
Achievability is unaffected because $\setE(\vec{y})$ is computed from $\vec{y}$,~$\mat{H}$, and~$\Nzero$, and is part of the COIL soft output that the decoder will receive (cf.\ \fref{defi:coil}).

%%%
\subsection{Proof of Corollary~\ref{cor:mwst}} 
\label{app:mwst}
In~\fref{eq:oracletree} and~\fref{eq:oracletreey}, the only term that depends on the tree is the sum of its edge weights. 
Thus, maximizing the rate means maximizing that sum; this is what a maximum-weight spanning tree does, and it proves statement~(i); for statement~(ii), the sum is maximized separately for each $\vec{y}$. 
The weights are MIs and thus nonnegative, so adding an edge never hurts and no forest can beat a spanning tree.
Statement~(iii) is~\fref{eq:treekl} combined with~\fref{eq:bicmloss}: since a KL divergence is nonnegative, the left-hand side of~\fref{eq:treekl} is at least zero, which is precisely~\fref{eq:treeceiling}, with equality when the divergence vanishes, i.e., when the posterior is the tree metric.

%%%
\subsection{Proof of Lemma~\ref{lem:boxplus}} 
\label{app:boxplus}
The message that reaches the check node from variable node~$x_k$ is $p(x_k)\propto \e^{\mu x_k/2}$, so the factor-to-variable message to $x_l$ is $m(x_l)=\sum_{x_k\in\{\pm1\}}\e^{\gamma_{k,l}x_kx_l/2}\,p(x_k)$, whose logit is
\begin{align}
  \ln\!\left(\frac{m(+1)}{m(-1)}\right)\!
  &= \ln\!\left(\frac{\e^{(\gamma_{k,l}+\mu)/2}+\e^{-(\gamma_{k,l}+\mu)/2}}
            {\e^{(\mu-\gamma_{k,l})/2}+\e^{(\gamma_{k,l}-\mu)/2}}\right)
   \\
  &= 2\tanh^{-1}\big(\tanh(\tfrac{\gamma_{k,l}}{2})\tanh(\tfrac{\mu}{2})\big),
  \label{eq:boxplusproof}
\end{align}
which equals $\gamma_{k,l}\boxplusop\mu$ by~\fref{eq:boxplus}. The same identity describes the output of a degree-$3$ check node whose third input is the constant logit $\gamma_{k,l}$~\cite{hagenauer96}.
Setting $\gamma_{k,l}=0$ gives $0\boxplusop\mu=0$.

%%%
\subsection{Proof of Lemma~\ref{lem:absorb}} 
\label{app:absorb}
By \fref{lem:ising}, every correction factor splits into two degree-$1$ factors $\e^{\delta_k^{(k,l)}x_k/2}$ and $\e^{\delta_l^{(k,l)}x_l/2}$ and the pure coupling factor $\e^{\gamma_{k,l}x_kx_l/2}$. 
A degree-$1$ factor sends the same constant message in every BP iteration, so associating it to a variable node is equivalent to adding $\delta_k^{(k,l)}$ to that node's channel input. 
Summing over all edges connected to bit $k$ gives~$\check{\lambda}_k$ of~\fref{eq:fieldcorrected}, and the coupling factors remain.
On edge $(k,l)$ at bit $k$, the transformed variable-to-factor logit is the original logit plus~$\delta_k^{(k,l)}$, and the transformed factor-to-variable logit is the original logit minus $\delta_k^{(k,l)}$.
These shifts preserve the BP updates and bit beliefs when applied to the initial messages as well.

%%%%%%%%%%%%%%%%%%%%%%%%%%%%%%%%%%%%%%%%%%%%%%%%%%%%%%%%%%%
%% Bibliography
%%%%%%%%%%%%%%%%%%%%%%%%%%%%%%%%%%%%%%%%%%%%%%%%%%%%%%%%%%%

\balance
\bstctlcite{IEEEexample:BSTcontrol} % keep repeated author names in the biography
\bibliographystyle{IEEEtran}
\bibliography{bib/publishers,bib/journals_proceedings_ect,bib/library,bib/idd}
\balance

\end{document}